\documentclass[11pt]{article}
\usepackage{hyperref}
\usepackage{booktabs}

\usepackage{graphicx}
\usepackage{epstopdf}
\usepackage{float} 
\usepackage{mathrsfs}
\usepackage{amsmath}
\usepackage{slashed}
\usepackage{bigints}
\usepackage{amsmath}
\usepackage{amsfonts}   
\usepackage{amssymb}
\usepackage{subfloat}

\begin{document}
\date{}
\title{{\bf{\Large Holographic $s$-wave condensation and Meissner-like effect in Gauss-Bonnet gravity with various non-linear corrections}}}

\author{
{\bf {\normalsize Shirsendu Dey}$^{}
$\thanks{shirsendu12@bose.res.in}},
 {\bf {\normalsize Arindam Lala}
$^{}$\thanks{arindam.physics1@gmail.com}}\\
$^{}$ {\normalsize  S.N. Bose National Centre for Basic Sciences,}\\{\normalsize JD Block, 
Sector III, Salt Lake, Kolkata 700098, India}\\
\\[0.3cm]
}
\date{}

\maketitle
\begin{abstract}
In this paper we have studied the onset of holographic $s$-wave condensate in the $(4+1)$ dimensional planar Gauss-Bonnet-AdS black hole background with several non-linear corrections to the gauge field. In the probe limit, performing explicit analytic computations, with and without magnetic field, we found that these higher order corrections indeed affect various quantities characterising the holographic superconductors. Also, performing a comparative study of the two non-linear electrodynamics it has been shown that the exponential electrodynamics has stronger effects on the formation of the scalar hair. We observe that our results agree well with those obtained numerically [Z. Zhao et. al., Nucl. Phys. B 871 [FS] (2013) 98].
\end{abstract}
\maketitle
\section{Introduction and motivations}
The AdS/CFT correspondence\cite{ref1}-\cite{ref2} has proven to be useful in describing several aspects of strongly coupled field theories from their weakly coupled dual gravity theories which lie in one higher dimension.\footnote{For excellent reviews on this correspondence see \cite{ref3}-\cite{ref6}.} Over the past several years this correspondence has been extensively used to explore certain field theoretic phenomena where conventional perturbation methods fail to give consistent results\cite{ref7}.

Among many sectors where this correspondence has been applied successfully, strongly interacting condensed matter theory has been central to the discussion where conventional perturbation techniques appear to be unfaithful. Using this holographic conjecture a holographic model of high $T_{c}$ superconductors has been proposed\cite{ref16}-\cite{ref17}. Further study of these phenomenological models reveal many of their interesting features which resemble that of the conventional superconductors\cite{ref18}.

Gravity theories in higher dimensions\footnote{For good reviews on gravity theories in higher dimensions see \cite{ref20}-\cite{ref20b}.} (greater than four) have earned repeated attentions in the past several decades with the advent of string theory. The primary motivation comes from the fact that the consistent description of string theory requires the inclusion of higher dimensional space-time. Certain aspects of string theory are well described by associating gravity theories with it. The effect of string theory on gravity may be studied by considering a low-energy effective action which describes classical gravity\cite{ref21}. This effective action must contain higher curvature terms and are needed to be ghost free\cite{ref22}. The Lovelock action is found to be consistent with these criteria\cite{ref23}.

Along with the conventional Maxwell electrodynamic theory non-linear electrodynamic theories (NED), which correspond to the higher derivative corrections to the Abelian gauge fields, have also become interesting topics of research for the past several decades. The primary motivation for introducing non-linear electrodynamic theory was to remove divergences in the self-energy of point-like charged particles\cite{ref38}. However, they have earned renewed attention over past several years since these theories naturally arise in the low-energy limit of the heterotic string theory\cite{ref24}-\cite{ref26}.

Besides the conventional Born-Infeld non-linear electrodynamics (BINE)\cite{ref38}, two new types of NEDs have been proposed recently, namely the exponential non-linear electrodynamics (ENE) and the logarithmic non-linear electrodynamics (LNE), in the context of static charged asymptotic black holes\cite{ref37,ref37a}. In fact, the matter actions with ENE and LNE yield the higher derivative corrections to the usual Maxwell action. On the other hand these NEDs possess many unique properties which are quite different from the Maxwell electrodynamics. For example, while solutions with LNE completely remove divergences in the electric fields at $r=0$, these divergences still remain in the solutions with ENE. But these divergences are much weaker than the usual Maxwell case\cite{ref37,ref37a}. Also, compared with Maxwell theory, solutions with LNE and ENE have different temperatures and electric potentials\cite{ref37,ref37a}. Another  novel property of these non-linear theories is that, their asymptotic black hole solutions are the same as that of a Reissner-Nordstr\"{o}m black hole\cite{ref37a}. On top of that, these types of non-linear theories retain some interesting properties (alike BINE) such as, absence of shock waves, birefringence etc.\cite{ref37,ref37a}. One further advantage of studying ENE and LNE over the Maxwell theory is that they provide an enriched platform to investigate generalized versions of NEDs in a systematic manner so as to reveal some general features of the effects of higher derivative corrections to the gauge fields in the theory concerned.

At this point of discussion it must be stressed that while gravity theories with NEDs give rise to many interesting gravity solutions which in many respects are different from the solutions with usual Maxwell electrodynamics\cite{ref27}-\cite{ref36}, these are also widely discussed in the context of gauge/string duality, specifically in the holographic study of condensed matter phenomena with holographic superconductors as specific examples. Holographic superconductors with non-linear electrodynamics\cite{ref45}-\cite{ref63} have been investigated alongside those with Maxwell electrodynamics\cite{ref18}. These studies show that the non-linearity in the theory indeed modifies the behaviors of the holographic condensates in non-trivial manners which cannot be observed in the conventional holographic superconductors with Maxwell electrodynamics. In other words the higher derivative corrections to the Abelian gauge fields are manifested as effects on certain properties of the dual holographic models. This observation motivates us to study models of holographic superconductors with ENE and LNE and look for the modifications they make on certain properties of the models compared to the Maxwell case. This also encourages us to make a comparative study between holographic models with different NEDs regarding their effects on the condensation formation. In this regard consideration of holographic models with ENE and LNE is another motivation of the present paper.

Along with the study of holographic superconductors in the framework of conventional Einstein gravity several  attempts have been made in order to study these objects in the presence of higher curvature corrections to the Einstein gravity in higher dimensions\cite{ref49, ref51, ref58}, \cite{ref61}-\cite{ref48}. These curvature corrections also modify some of the properties of the holographic superconductors, such as $(i)$ the observed constancy of the ratio of the frequency gap of the real part of the conductivity to the critical temperature of the superconductor\cite{ref41} breaks down for Gauss-Bonnet superconductors\cite{ref39}, \cite{ref39a}-\cite{ref44a}, $(ii)$ in certain generalized cases of different values of the Gauss-Bonnet correction changes the order of the phase transition\cite{ref44a}, $(iii)$ the ratio of the shear viscosity to the entropy density ($\eta/s\geq 1/4\pi$) in CFT dual to the Einstein-Gauss-Bonnet gravity changes significantly with the Gauss-Bonnet coupling\cite{ref63a, ref63b}.

Apart from these, the effects of external magnetic fields on the holographic superconductors with or without these nontrivial non-linear corrections have been studied. These studies reveal several interesting properties of these superconductors which resemble certain properties of conventional superconductors, such as the Meissner effect, vortex and droplet solutions etc.\cite{ref54,ref55},~\cite{ref64}-\cite{ref69a}.

It must be emphasized that holographic superconductors with several NEDs (BINE, ENE, LNE) has been studied in Ref.\cite{ref59} in the planar Schwarzschild-AdS black hole background without taking into account higher curvature corrections to the Einstein gravity. Also, in Ref.\cite{ref63} a holographic model with BINE in the Gauss-Bonnett gravity has been considered. Surprisingly most of the computations have been performed using numerical methods. Therefore, it would be nice to perform analytic analysis of holographic superconductor models with non-linear electrodynamic fields (ENE and LNE) in the Gauss-Bonnett black hole background.

Considering all the above mentioned facts, in this paper we have made an extensive analytic investigation of the holographic model of superconductors mentioned at the end of the previous paragraph in the presence of an external magnetic field. The primary motivations of our study may be summarized as follows: (i) investigating the effects of higher curvature as well as higher derivative corrections on the properties of holographic superconductors, (ii) exploring the effects of an external magnetic field on the holographic condensates and determining how the non-linear corrections modify the critical value of the magnetic field, (iii) making a comparison between the Maxwell theory and the non-linear electrodynamic theories regarding their effects on the formation of the scalar condensates, (iv) comparing the non-linear electrodynamic theories in an attempt to see which one has stronger effects on the formation of the scalar condensates.

The present paper has been organized as follows. In section 2, the basic setup for $s$-wave holographic superconductor with two different non-linear electrodynamics (ENE and LNE) in the planar $(4+1)$-dimensional Gauss-Bonnet AdS black hole background is given. In section 3, we have calculated various properties of the $s$-wave holographic superconductor with exponential electrodynamics (ENE) which include the critical temperatures for condensation and the expectation values of the condensation operator in the absence of external magnetic field. In section 4, we have discussed the effects of an external static magnetic field on this holographic superconductor and calculated the critical magnetic field for condensation. We have drawn our conclusions and discussed some of the future scopes in section 5. Finally, in appendix A we have given the expressions of the aforementioned quantities for the $s$-wave holographic superconductor with logarithmic electrodynamics (LNE) and derived a necessary mathematical relation.
\label{sec:intro}
\section{Basic set up}
The effective action for the higher curvature Lovelock gravity in an arbitrary dimension $d$ may be written as\cite{ref23},
\begin{equation}
\label{eq:2.1}
\mathcal{S}_{grav}=\dfrac{1}{16\pi G_{d}}\int d^{d}x \sqrt{-g}\sum_{i=0}^{[d/2]}\alpha_{i}\mathcal{L}_{i}
\end{equation}
where, $\alpha_{i}$ is an arbitrary constant, $\mathcal{L}_{i}$ is the \textit{Euler density} of a $2i$ dimensional manifold and $G_{d}$ is the Gravitational constant in $d$-dimensions. In our subsequent analysis we shall consider the coordinate system where $G_{d}=\hbar=k_{B}=c=1$.

In this paper we shall be mainly concerned with the $(4+1)$-dimensional Einstein-Gauss-Bonnet gravity in anti-de Sitter (AdS) space. Thus, the effective action \eqref{eq:2.1} can be written as,
\begin{eqnarray}
 \label{eq:2.2}
\mathcal{S}_{grav}&=&\frac{1}{16\pi}\int d^{5}x \sqrt{-g}\,\Big(\alpha_{0}\mathcal{L}_{0} +\alpha_{1}\mathcal{L}_{1}+\alpha_{2}\mathcal{L}_{2}\Big)\nonumber\\
&=&\frac{1}{16\pi}\int d^{5}x \sqrt{-g}\,\Big(-2 \Lambda +\mathcal{R}+\alpha\mathcal{L}_{2}\Big)
\end{eqnarray}
where $ \Lambda $ is the cosmological constant given by $ -6/l^2 $, $ l $ being the AdS length, $ \alpha_{2}\equiv \alpha $ is the Gauss-Bonnet coefficient, $\mathcal{L}_{1}= \mathcal{R} $ is the usual Einstein-Hilbert Lagrangian and $ \mathcal{L}_{2}=R_{\mu\nu\gamma\delta}R^{\mu\nu\gamma\delta}
-4R_{\mu\nu}R^{\mu\nu}+\mathcal{R}^{2}$ is the Gauss-Bonnet Lagrangian.

The Ricci flat solution for the action \eqref{eq:2.2} is given by\cite{ref39,ref71,ref72}
\begin{equation}
\label{eq:2.3}
ds^{2}=-f(r)dt^{2}+f(r)^{-1}dr^{2}+r^{2}(dx^{2}+dy^{2}+dz^{2})
\end{equation}
where the metric function is\footnote{Without loss of generality, we can choose $l=1$, which follows from the scaling properties of the equation of motion.}\cite{ref39,ref71,ref72},
\begin{equation}
\label{eq:2.4}
f(r)=\frac{r^2}{2a}\left( 1-\sqrt{1-4a\left(1-\frac{M}{r^4}\right)}\right).
\end{equation}

In \eqref{eq:2.4}, $M$ is the mass of the black hole which may be expressed in terms of the horizon radius ($r_{+}$) as, $r_{+}=M^{1/4}$\cite{ref51,ref39,ref44}; the parameter $a$ is related to the coefficient $\alpha$ as, $a=2\alpha$. It is to be noted that, in order to avoid naked singularity we must have $a\leq 1/4$\cite{ref51,ref39}, whereas, considering the causality of dual field theory on the boundary the lower bound of $a$ is found to be $a\geq -7/36$\cite{refnew}. Also, in the asymptotic infinity  ($r\rightarrow\infty$) we may write the metric function \eqref{eq:2.4} as,
\begin{equation}
\label{eq:2.6}
f(r)\sim \frac{r^{2}}{2a}\Big(1-\sqrt{1-4a}\Big).
\end{equation}

Thus, the effective AdS radius can be defined as\cite{ref39},
\begin{equation}
\label{eq:2.7}
L_{eff}^{2}=\dfrac{2a}{1-\sqrt{1-4a}}.
\end{equation}

Note that, in the limit $a\rightarrow1/4$, $L_{eff}^{2}=0.5$. This limit is known as the Chern-Simons limit\cite{ref73}. 

The Hawking temperature of the black hole may be obtained by analytic continuation of the metric at the horizon ($r_{+}$) and is given by\cite{ref51,ref39,ref44},
\begin{equation}
\label{eq:2.5}
T=\dfrac{r_{+}}{\pi}.
\end{equation}

In this paper we shall study the $s$-wave holographic superconductor in the framework of various non-linear electrodynamics in the $(4+1)$-dimensional planar Gauss-Bonnet AdS (GBAdS) black hole background. For this purpose, we shall consider a matter Lagrangian which consists of a charged $U(1)$ gauge field, $A_{\mu}$, and a charged massive complex scalar field, $\psi$. Thus, the matter action for the theory may be written as\cite{ref16,ref17},
\begin{equation}
\label{eq:2.8}
\mathcal{S}_{matter}=\int d^{5}x\sqrt{-g}\Big(L(F)-|\nabla_{\mu}\psi- iA_{\mu}\psi |^{2}-m^{2}\psi^{2}\Big)
\end{equation}
where $m$ is the mass of the scalar field. Moreover, we shall carry out all the calculations in the probe limit\cite{ref18}. In this limit, gravity and matter decouple and the backreaction of the matter fields (scalar field and gauge field) can be suppressed in the neutral AdS black hole background. This is achieved by rescaling the matter action \eqref{eq:2.8} by the charge, $q$, of the scalar field and then considering the limit $q\rightarrow\infty$. The generosity of this approach is that we can simplify the problems without hindering the physical properties of the system. The term $L(F)$ in \eqref{eq:2.8} corresponds to the Lagrangian for the non-linear electrodynamic field. In different non-linear theories the Lagrangian $L(F)$ can take the following forms\cite{ref37,ref59}:
\begin{eqnarray}
\label{eq:2.9}
 L(F)=
\left\{
\begin{array}{lr}
\dfrac{1}{4b}\Big(e^{-bF^{\mu\nu}F_{\mu\nu}}-1\Big),  &\text{for ENE} \\ \\
\dfrac{-2}{b}~ln\Big(1+\frac{1}{8}bF^{\mu\nu}F_{\mu\nu}\Big), &\text{for LNE}                            
\end{array}
\right.
\end{eqnarray}
Note that in the limit $b\rightarrow 0$, we recover the usual Maxwell term: $L(F)|_{b\rightarrow 0}=-\frac{1}{4}F^{\mu\nu}F_{\mu\nu}$.


In order to solve the equations of motion resulting from the variation of the action \eqref{eq:2.8} w.r.to the gauge and scalar fields we shall choose the following ansatz for the two fields concerned\cite{ref54}:
\begin{subequations}\label{eq:2.10}
\begin{align}
\label{eq:2.10:1}
A_{\mu} =&\Big(\phi(r),0,0,0,0\Big),
\\
\label{eq:2.10:2}
\psi=&\psi(r).
\end{align}
\end{subequations}
It is to be noted that, the above choices of the fields are justified since it is seen that under the transformations $A_{\mu}\rightarrow A_{\mu}+\partial_{\mu}\theta$ and $\psi\rightarrow\psi e^{i\theta}$ the above equations of motion remain invariant. This demands that the phase of $\psi$ remains constant and we may take $\psi$ to be real without any loss of generality. 

With the change of coordinates $z=\frac{r_{+}}{r}$, where the horizon ($r=r_{+}$) is at $z=1$ and the boundary ($r\rightarrow \infty$) is at $z=0$, the equations of motion for the the scalar field ($\psi(z)$) and the $U(1)$ gauge field ($A_{\mu}$) may be obtained as\cite{ref39,ref44},
\begin{eqnarray}
\label{eq:2.11}
\psi''(z)+\left(\frac{f'(z)}{f(z)}-\frac{1}{z}\right)\psi'(z)
+\frac{\phi^{2}(z)\psi(z) r_+^{2}}{z^{4}f^{2}(z)}-\frac{m^2\psi(z) r_+^{2}}{z^{4}f(z)}=0,
\end{eqnarray}
\begin{eqnarray}
\label{eq:2.13}
\Bigg(1+\dfrac{4bz^{4}\phi'^{2}(z)}{r_{+}^{2}}\Bigg)\phi''(z) -\frac{1}{z}\phi'(z)+\dfrac{8bz^3}{r_+^2}\phi'^{3}(z)-\frac{2\psi^{2}(z)\phi(z){r_+^2}}{f(z)z^4}e^{-2bz^{4}\phi'^{2}(z)/r_{+}^{2}}=0,\qquad\text{for ENE}
\end{eqnarray}

\begin{eqnarray}
\label{eq:2.14}
\Bigg(1+\dfrac{bz^{4}\phi'^{2}(z)}{4r_{+}^{2}}\Bigg)\phi''(z) -\frac{\phi'(z)}{z}+\frac{5bz^3}{4r_+^2}\phi'^{3}(z)-\frac{2\psi^{2}(z)\phi(z){r_+^2}}{f(z)z^4}\Bigg(1-\dfrac{bz^{4}\phi'^{2}(z)}{4r_{+}^{2}}\Bigg)^{2}=0,\qquad\text{for LNE}.
\end{eqnarray}

In order to solve the above set of equations we shall choose the following boundary conditions: 

 (i) At the horizon ($z=1$) one must have, for $m^2=-3$,\footnote{For the rest of the analysis of our paper we choose $m^{2}=-3$. This ensures that we are above the Breitenlohner-Freedman bound\cite{ref74,ref75}.}
\begin{equation}
\label{eq:2.15}
\phi (1)=0,~~~\psi^{'}(1)=\frac{3}{4}\psi(1) 
\end{equation}

(ii) In the asymptotic AdS region ($ z\rightarrow 0 $) the solutions for the scalar potential and the scalar field may be expressed as,
\begin{subequations}\label{eq:2.16}
\begin{align}
\label{eq:2.16:1}
\phi(z) =& ~\mu - \frac{\rho}{r_{+}^{2}}z^{2},
\\
\label{eq:2.16:2}
\psi(z)=& 
~D_{-}z^{\lambda_{-}}+D_{+}z^{\lambda_{+}}
\end{align}
\end{subequations}
where $\lambda_{\pm} = 2\pm\sqrt{4-3L_{eff}^{2}}$ is the conformal dimension of the condensation operator $\mathcal{O}_{i}$ ($i=1,2$) in the boundary field theory, $\mu$ and $\rho$ are identified as the chemical potential and the charge density of the dual field theory, respectively. It is interesting to note that, since we have considered $m^{2}<0$ in our analysis, we are left with the two different condensation operators of different dimensionality corresponding to the choice of quantization of the scalar field $\psi$ in the bulk. We choose $D_{-}=0$. Then, according to the AdS/CFT correspondence $D_{+}\equiv\langle\mathcal{O}_{2}\rangle$, the expectation value of the condensation operator in the dual field theory.
\label{sec:setup}
\section{$s$-wave condensate and its critical behaviour without magnetic field}
In this section we shall analytically derive the critical temperature for condensation, $T_{c}$, for the holographic $s$-wave condensate with two types of non-linear electrodynamics mentioned in the previous section. As a next step, we shall determine the normalized condensation operator and the critical exponent associated with the condensation values in the presence of these non-linear theories in the background of $(4+1)$-dimensional Gauss-Bonnet AdS black hole. In this way we would be able to demonstrate the effects of the Gauss-Bonnet coupling coefficient ($a$) as well as non-linear parameter ($b$) on these condensates.

In order to carry out our analysis we have adopted a well known analytic technique which is known as the matching method\cite{ref39}. In this method, we first determine the leading order solutions of the equations of motion \eqref{eq:2.11}, \eqref{eq:2.13}, \eqref{eq:2.14} near the horizon ($1\geq z>z_{m}$) and at the asymptotic infinity ($z_{m}>z\geq 0$) and then match these solutions smoothly at the intermediate point, $z_{m}$.\footnote{In this paper we pursue our analytic investigation in the same spirit as in Refs.\cite{ref55},~\cite{ref39} and match the leading order solutions near the horizon and the boundary at the intermediate point $z_{m}=0.5$. It may be stressed that the qualitative features of the analytical approximation does not change for other values of $z_{m}$ ($0< z_{m}\leq 1$) and differences in the numerical values are not too large\cite{ref39}. Therefore throughout our analysis we shall choose $z_{m}=0.5$ while obtaining numerical values and plotting various quantities. In fact, as mentioned in Ref.\cite{Kanno}, this choice of the matching point roughly corresponds to the geometrical mean of the horizon radius and the AdS scale. Interestingly, with this choice of $z_{m}$ our results are fairly consistent with Ref.\cite{ref39} for $b=0$.} It is to be noted that the matching method helps us to determine the values of the critical temperature as well as of condensation operator only approximately, in the leading order of the non-linear parameter, $b$. Moreover, this method provides us a much better understanding of the effects the Gauss-Bonnet coefficient ($a$) as far as analytic computation is concerned\cite{ref39}.

In this section we shall perform the analysis for the holographic $s$-wave superconductor with exponential electrodynamics only. Since, the analysis for the holographic superconductor with logarithmic electrodynamics closely resemblances to that of the previous one, we shall only present the results corresponding to this model in appendix A.1.

Let us first consider the solutions of the gauge field, $\phi(z)$, and the scalar field, $\psi(z)$, near the horizon ($z=1$). We Taylor expand both $\phi(z)$ and $\psi(z)$ near the horizon as\cite{ref39},
\begin{equation}
\label{eq:3.1}
\phi (z)=\phi (1) - \phi^{'}(1)(1-z) + \frac{1}{2}\phi^{''}(1)(1-z)^{2} + \cdots
\end{equation}
\begin{equation}
\label{eq:3.2}
\psi(z)=\psi(1)-\psi^{'}(1)(1-z)+\frac{1}{2} \psi^{''}(1)(1-z)^{2} + \cdots
\end{equation}

It is to be noted that, in \eqref{eq:3.1} and \eqref{eq:3.2}, we have considered $\phi'(1)<0$ and $\psi(1)>0$ in order to make $\phi(z)$ and $\psi(z)$ positive. This can be done without any loss of generality.

Near the horizon, $z=1$, we may write from \eqref{eq:2.11}
\begin{equation}
\label{eq:3.3}
\psi^{''}(1)=\left[\frac{1}{z}\psi^{'}(z) \right]_{z=1}-\left[\frac{f^{'}(z)\psi^{'}(z)}{f(z)} \right]_{z=1}-\left[\frac{{\phi}^2(z)\psi(z)r_+^2}{z^4f^2(z)}\right]_{z=1}-\left[\frac{3\psi(z){r_+}^2}{z^4f(z)}\right]_{z=1}.
\end{equation}

Using the L'H\^{o}pital's rule and the values $f^{'}(1)=-4r_{+}^2$, $f^{''}(1)=4{r_+}^2+32a{r_+}^2$, we may express \eqref{eq:3.3} in the following form:
\begin{equation}
\label{eq:3.4}
\psi^{''}(1)=-\frac{5}{4}\psi^{'}(1)+8a\psi^{'}(1)-\frac{{\phi^{'}}^2(1)\psi(1)}{16{r_+}^2}.
\end{equation}

Finally, using the boundary condition \eqref{eq:2.15}, the Taylor expansion \eqref{eq:3.2} can be rewritten as,
\begin{equation}
\label{eq:3.5}
\psi(z)=\frac{1}{4}\psi(1)+\frac{3}{4}\psi(1)z +(1-z)^2\left[-\frac{15}{64}+\frac{3a}{2}-\frac{\phi'^{2}(1)}{64{r_+}^2}\right]\psi(1).
\end{equation}

Near the horizon, $z=1$, from \eqref{eq:2.13} we may write
\begin{equation}
\label{eq:3.6}
\phi^{''}(1)= \dfrac{1}{\left(1+\frac{4b}{r_{+}^{2}}\phi'^{2}(1)\right)}\left[\phi^{'}(1)-\frac{8b}{{r_+}^2}\phi^{'3}(1)-\frac{\psi^{2}(1)\phi^{'}(1)}{2}e^{-2b\phi'^{2}(z)/r_{+}^{2}}\right].
\end{equation}

In obtaining \eqref{eq:3.6} we have considered that the metric function, $f(z)$, can also be Taylor expanded as in \eqref{eq:3.1}, \eqref{eq:3.2}.

Substituting \eqref{eq:3.6} in \eqref{eq:3.1} and using \eqref{eq:2.15} we finally obtain,
\begin{equation}
\label{eq:3.7}
\phi(z)= -\phi^{'}(1){(1-z)}+\frac{1}{2}{(1-z)^2}\left[1-\frac{8b}{{r_+}^2}\phi'^{2}(1)-\frac{\psi^{2}(1)}{2}e^{-2b\phi'^{2}(z)/r_{+}^{2}}\right]\dfrac{\phi'(1)}{\left(1+\frac{4b}{r_{+}^{2}}\phi'^{2}(1)\right)}.
\end{equation}  

Now, using the method prescribed by the matching technique\cite{ref39}, we match the solutions \eqref{eq:2.16:1}, \eqref{eq:2.16:2}, \eqref{eq:3.5} and \eqref{eq:3.7} at the intermediate point $z=z_{m}$. It is very much evident that the matching of the two asymptotic solutions smoothly at $z=z_{m}$ requires the following four conditions:
\begin{equation}
\label{eq:3.8}
\mu-\frac{\rho z_{m}^{2}}{r_{+}^{2}}=\beta(1-z_{m})-\frac{\beta}{2}(1-z_{m})^{2}\left[\dfrac{1-8b\tilde{\beta}^{2}}{1+4b\tilde{\beta}^{2}}-\frac{\alpha^{2}}{2}\frac{e^{-2b\tilde{\beta}^{2}}}{1+4b\tilde{\beta}^{2}}\right]
\end{equation}

\begin{equation}
\label{eq:3.9}
-\frac{2\rho z_{m}}{r_{+}^{2}}=-\beta+\beta(1-z_{m})\left[\dfrac{1-8b\tilde{\beta}^{2}}{1+4b\tilde{\beta}^{2}}-\frac{\alpha^{2}}{2}\frac{e^{-2b\tilde{\beta}^{2}}}{1+4b\tilde{\beta}^{2}}\right]
\end{equation}

\begin{equation}
\label{eq:3.10}
D_{+}z_{m}^{\lambda_{+}}=\frac{\alpha}{4}+\frac{3\alpha z_{m}}{4}+\alpha (1-z_{m})^{2}\left[\frac{-15}{64}+\frac{3a}{2}-\frac{\tilde{\beta}^{2}}{64}\right]
\end{equation}

\begin{equation}
\label{eq:3.11}
\lambda_{+}D_{+}z_{m}^{\lambda_{+}}=\frac{3\alpha z_{m}}{4}-2\alpha z_{m}(1-z_{m})\left[\frac{-15}{64}+\frac{3a}{2}-\frac{\tilde{\beta}^{2}}{64}\right]
\end{equation}
where we have set $\psi(1)=\alpha$, $-\phi'(1)=\beta$ ($\alpha,\beta >0$), $\tilde{\beta}=\frac{\beta}{r_{+}}$ and $D_{-}=0$ [cf.\eqref{eq:2.16:2}].

From \eqref{eq:3.9}, using \eqref{eq:2.5}, we obtain,
\begin{equation}
\label{eq:3.12}
\alpha^{2}=\dfrac{2z_{m}}{(1-z_{m})}e^{2b\tilde{\beta}^{2}}\left(1+\frac{4b\tilde{\beta}^{2}(3-2z_{m})}{z_{m}}\right)\left(\dfrac{T_{c}}{T}\right)^{3}\left(1-\dfrac{T^{3}}{T_{c}^{3}}\right).
\end{equation}

Since, in our entire analysis we intend to keep terms which are only linear in the non-linear parameter, $b$, \eqref{eq:3.12} can be approximated as,
\begin{equation}
\label{eq:3.13}
\alpha^{2}\approx\dfrac{2z_{m}}{(1-z_{m})}\left(1+\frac{6b\tilde{\beta}^{2}(2-z_{m})}{z_{m}}\right)\left(\dfrac{T_{c}}{T}\right)^{3}\left(1-\dfrac{T^{3}}{T_{c}^{3}}\right).
\end{equation}

Here the quantity $T_{c}$ may be identified as the critical temperature for condensation and is given by,
\begin{equation}
\label{eq:3.14}
T_{c}=\left[\dfrac{2\rho}{\tilde{\beta}\pi^{3}}\left(1-\frac{12b\tilde{\beta}^{2}(1-z_{m})}{z_{m}}\right)\right]^{\frac{1}{3}}.
\end{equation}

Now from \eqref{eq:3.10} and \eqref{eq:3.11} we obtain,
\begin{equation}
\label{eq:3.15}
D_{+}=\dfrac{(3z_{m}^{2}+5z_{m})}{4\left(\lambda_{+}+(2-\lambda_{+})z_{m}\right)}\left(\frac{1}{z_{m}}\right)^{\lambda_{+}}\alpha,
\end{equation}
\begin{equation}
\label{eq:3.16}
\tilde{\beta}=8\left[\frac{-15}{64}+\frac{3a}{2}-\dfrac{(3z_{m}^{2}+5z_{m})(1+\lambda_{+})}{4\left(\lambda_{+}+(2-\lambda_{+})z_{m}\right)\left(1-4z_{m}+3z_{m}^{2}\right)}
+\dfrac{(1+6z_{m})}{4\left(1-4z_{m}+3z_{m}^{2}\right)}\right]^{\frac{1}{2}}.
\end{equation}

Finally, using \eqref{eq:2.5}, \eqref{eq:3.13} and \eqref{eq:3.15}, near the critical temperature, $T\sim T_{c}$, we may write the expectation value, $\langle
\mathcal{O}_{2}\rangle$, of the condensation operator in the following form\footnote{Here we have used the relation $(1-t^{3})=(1-t)(1+t+t^{2})$ for any arbitrary variable $t$.}: 
\begin{equation}
\label{eq:3.17}
\dfrac{\langle
\mathcal{O}_{2}\rangle^{\frac{1}{\lambda_{+}}}}{T_{c}} =\left(\frac{\pi}{z_{m}}\right)\left[\dfrac{(3z_{m}^{2}+5z_{m})}{4\left(\lambda_{+}+(2-\lambda_{+})z_{m}\right)}\right]^{\frac{1}{\lambda_{+}}}
\left[\dfrac{6z_{m}}{(1-z_{m})}\left(1+\frac{6b\tilde{\beta}^{2}(2-z_{m})}{z_{m}}\right)\left(1-\dfrac{T}{T_{c}}\right)\right]^{\frac{1}{2\lambda_{+}}}.
\end{equation}
In \eqref{eq:3.17} we have normalized $\langle\mathcal{O}_{2}\rangle$ by the critical temperature, $T_{c}$, to obtain a dimensionless quantity\cite{ref39}.
In the similar fashion we can also calculate the critical temperature, $T_{c}$, and condensation operator, $\langle\mathcal{O}_{2}\rangle$, for the holographic superconductors with LNE. The corresponding expressions for the above mentioned quantities are given in Appendix~\ref{LNE}.

\begin{figure}[h]
\centering 
\includegraphics[width=.40\textwidth,trim=0 0 0 0,clip]{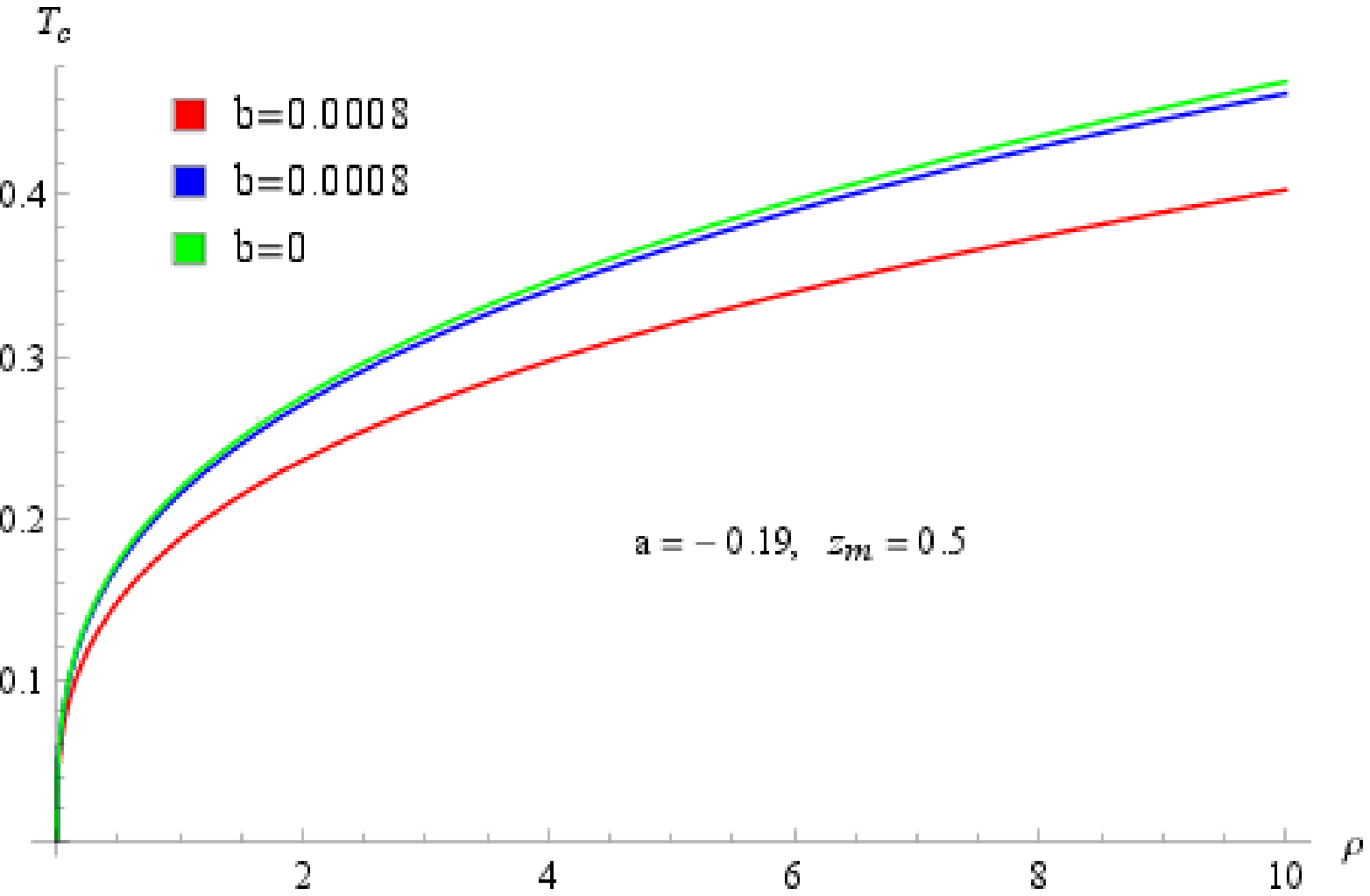}
\hfill
\includegraphics[width=.40\textwidth,origin=c,angle=0,]{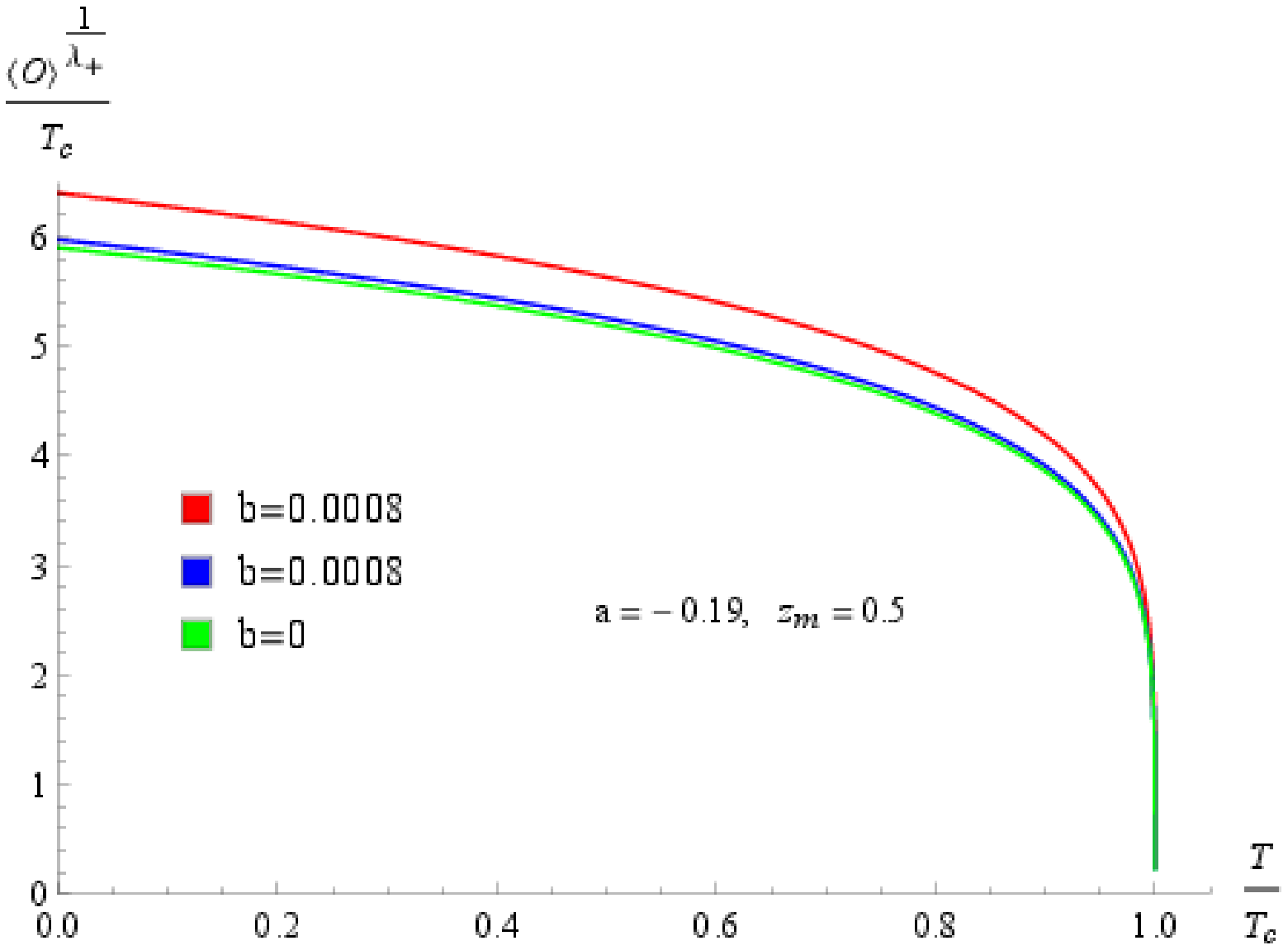}
\vfill
\includegraphics[width=.40\textwidth,trim=0 0 0 0,clip]{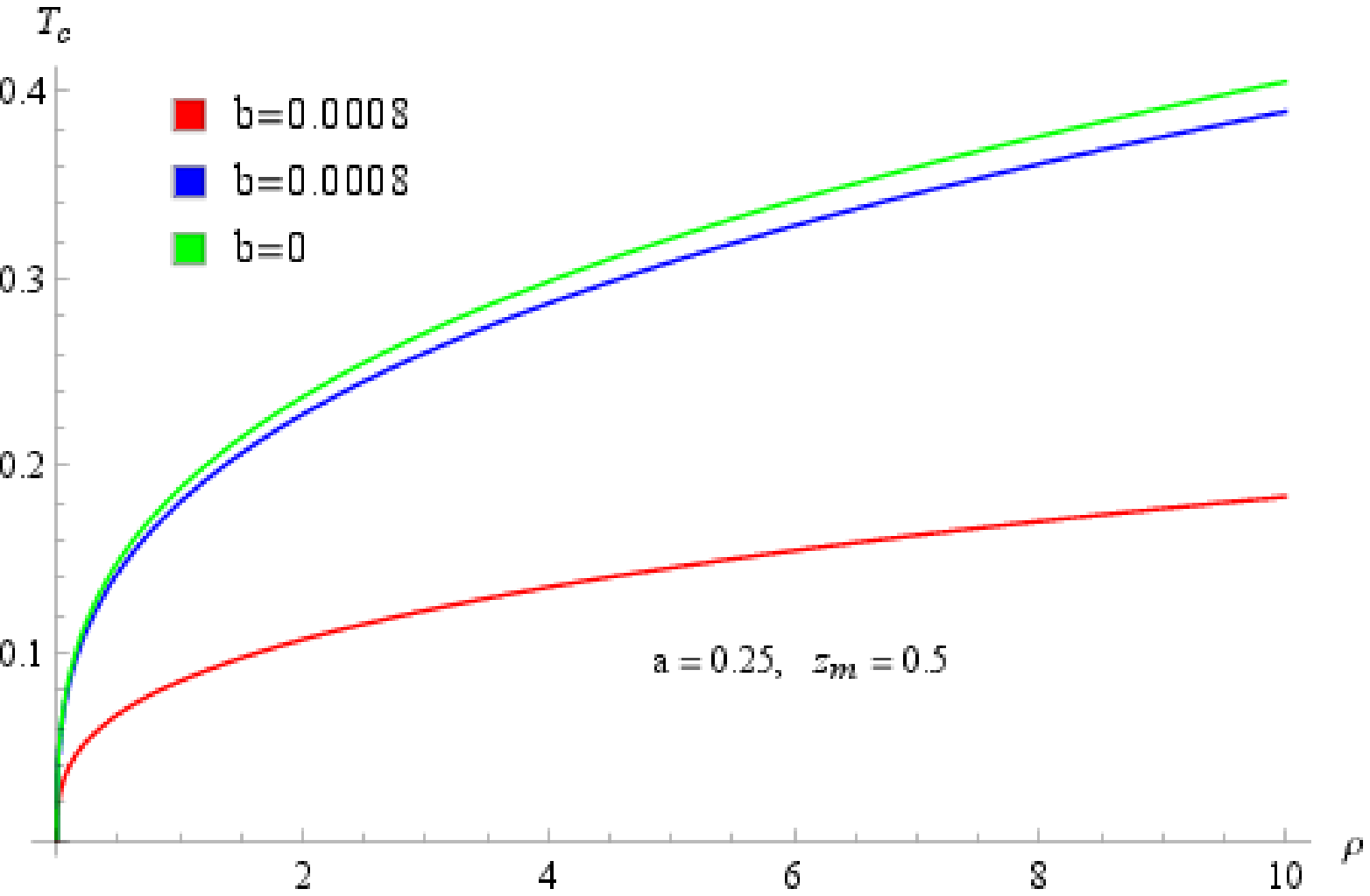}
\hfill
\includegraphics[width=.40\textwidth,origin=c,angle=0,]{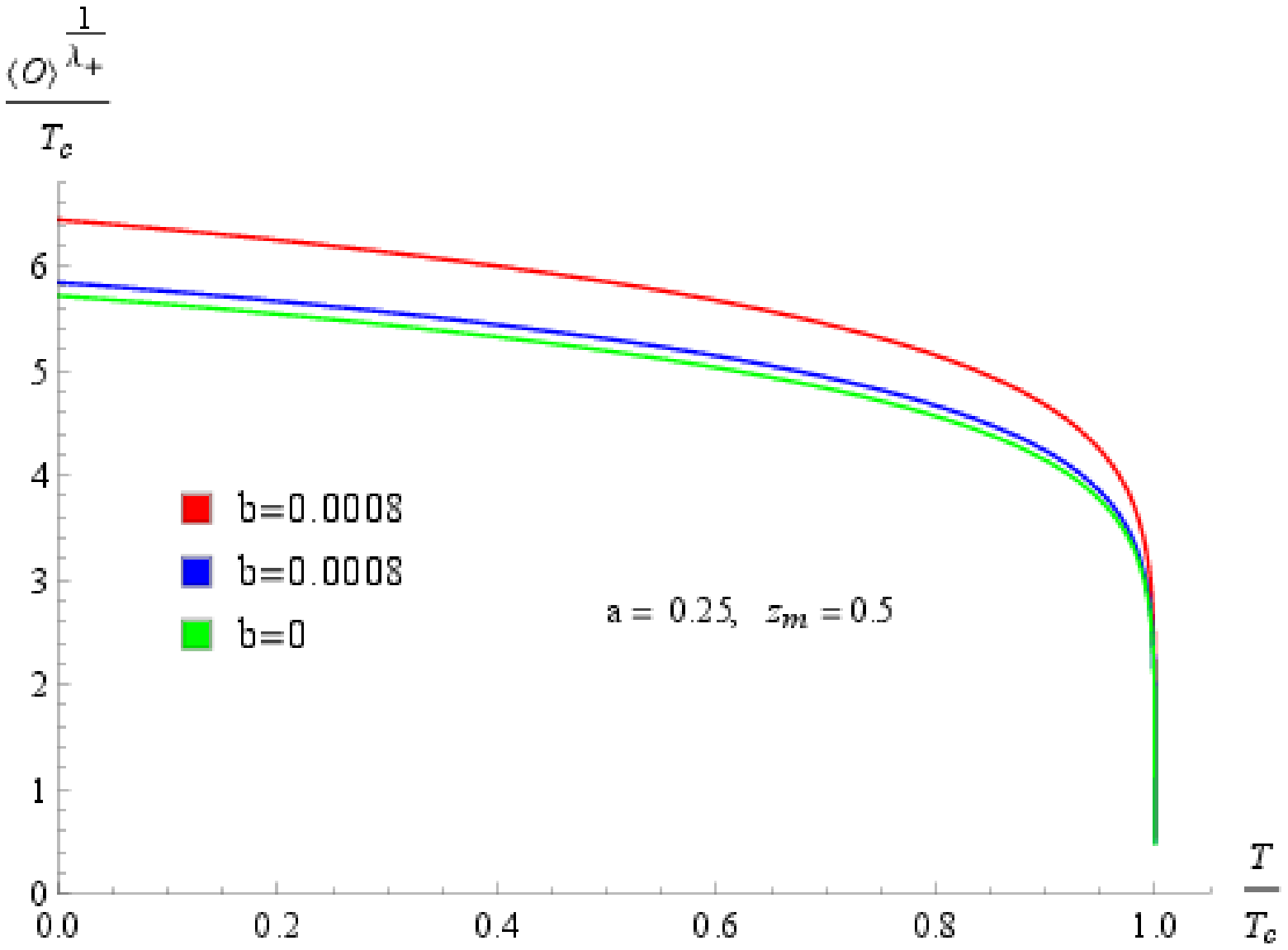}
\caption{\label{fig:1} Plots of critical temperature ($T_{c}-\rho$) and normalized condensation operator ($\langle\mathcal{O}_{2}\rangle^{\frac{1}{\lambda_{+}}}/T_{c}-T/T_{c}$) for different electrodynamic theories. The green, blue and red curves correspond to Maxwell, LNE and ENE, respectively.}
\end{figure}
In the next section we shall be mainly concerned with the effects of magnetic field on this $s$-wave holographic superconductor with the two different types of non-linear electrodynamics mentioned earlier. But, before that we would like to make some comments on the results obtained so far. These may be put as follows:

\hspace*{1cm}(i) From \eqref{eq:3.14} (and \eqref{eq:A1}) it is evident that in order to have a meaningful notion of the critical temperature, $T_{c}$, there must have an upper bound to the non-linear coupling parameter, $b$. The upper bounds corresponding to two non-linear theories are given below:
\begin{eqnarray}
\label{eq:3.22}
 b\leq
\left\{
\begin{array}{lr}
\dfrac{z_{m}(\lambda_{+}+2z_{m}-\lambda_{+}z_{m})}{12(1+96a\lambda_{+}-6(32a-13)(\lambda_{+}-1)z_{m}+3(32a-5)(\lambda_{+}-2)z_{m}^{2})},  &\text{for ENE} \\ \\
\dfrac{2z_{m}(\lambda_{+}+2z_{m}-\lambda_{+}z_{m})}{3(1+96a\lambda_{+})-18(32a-13)(\lambda_{+}-1)z_{m}+9(32a-5)(\lambda_{+}-2)z_{m}^{2}}, &\text{for LNE}                            
\end{array}
\right.
\end{eqnarray}

Note that, with our choice $z_{m}=0.5$ and for fixed values of $a$, this upper bound is smaller for ENE compared to LNE.

\hspace*{1cm}(ii) The critical temperature, $T_{c}$, decreases as we increase the values of the non-linear parameter, $b$ (Table:\ref{tab:1}). This feature is general for the two types of holographic superconductors considered in this paper. It must be remarked that, without any non-linear corrections ($b=0$) the critical temperature is larger than the above two cases. For example, $T_{c}=0.1907\rho^{1/3}$ for $a=0.2$, $z_{m}=0.5$. This suggests the onset of a harder condensation. Another nontrivial and perhaps the most interesting feature of our present analysis is that, for a particular value of the non-linear parameter, $b$, the value of the critical temperature, $T_{c}$, for the holographic condensate with ENE is smaller than that with LNE (Table:\ref{tab:1}) showing stronger effects of the former on the condensation. It is also noteworthy that similar feature was obtained numerically by the authors of \cite{ref59} in the planar Schwarzschild-AdS black hole background. 

\hspace*{1cm}(iii) The condensation gap for the holographic condensate with non-linear electrodynamics is more than that with Maxwell electrodynamics (Figure:\ref{fig:1}). On top of that, holographic superconductors with ENE exhibit larger gap compared with that with LNE. This suggests that the formation of the scalar hair is more difficult for the holographic condensate with ENE\cite{ref59}.

\begin{figure}[h]
\centering 
\includegraphics[width=.45\textwidth,trim=0 0 0 0,clip]{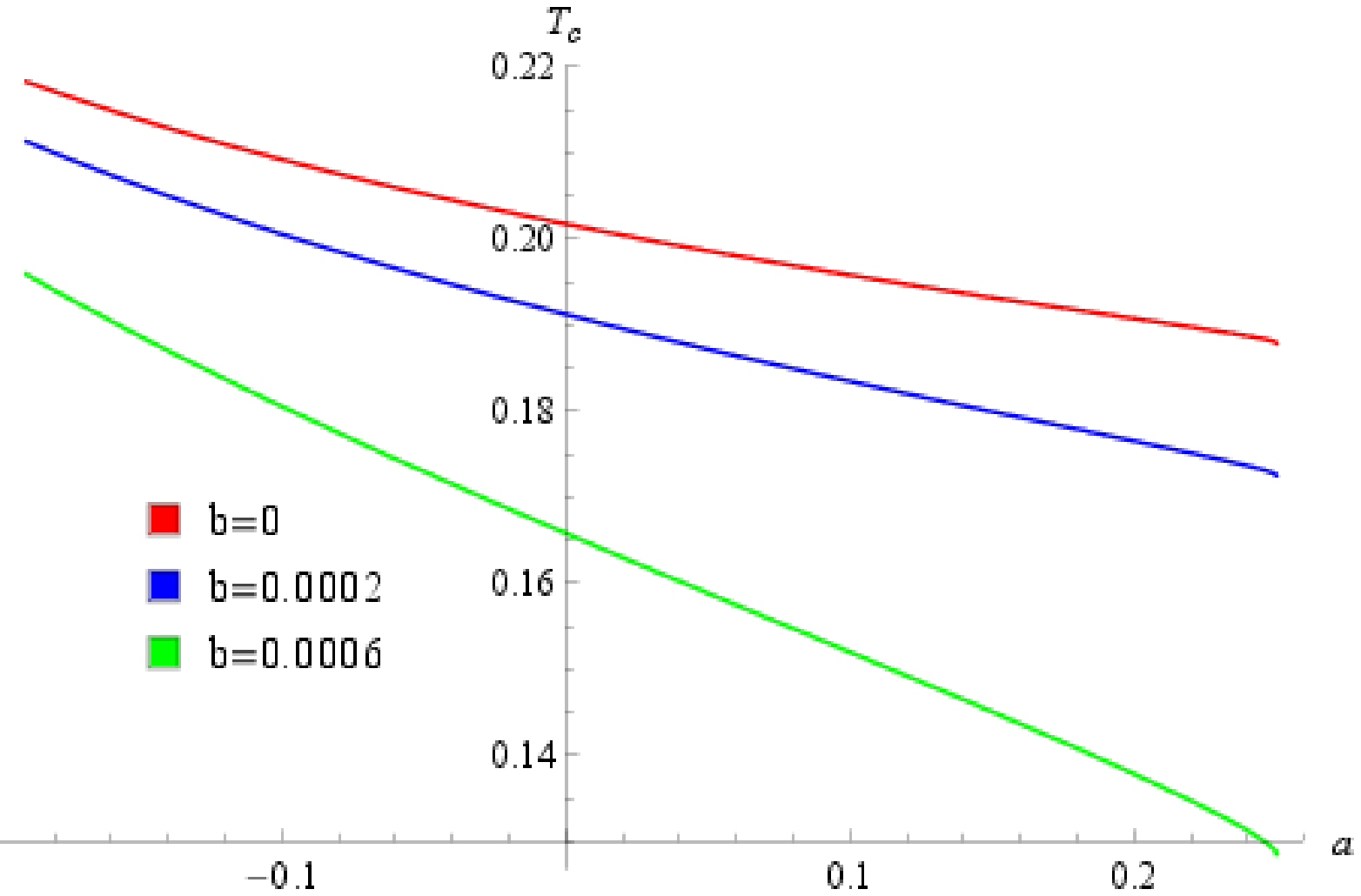}
\hfill
\includegraphics[width=.45\textwidth,origin=c,angle=0,]{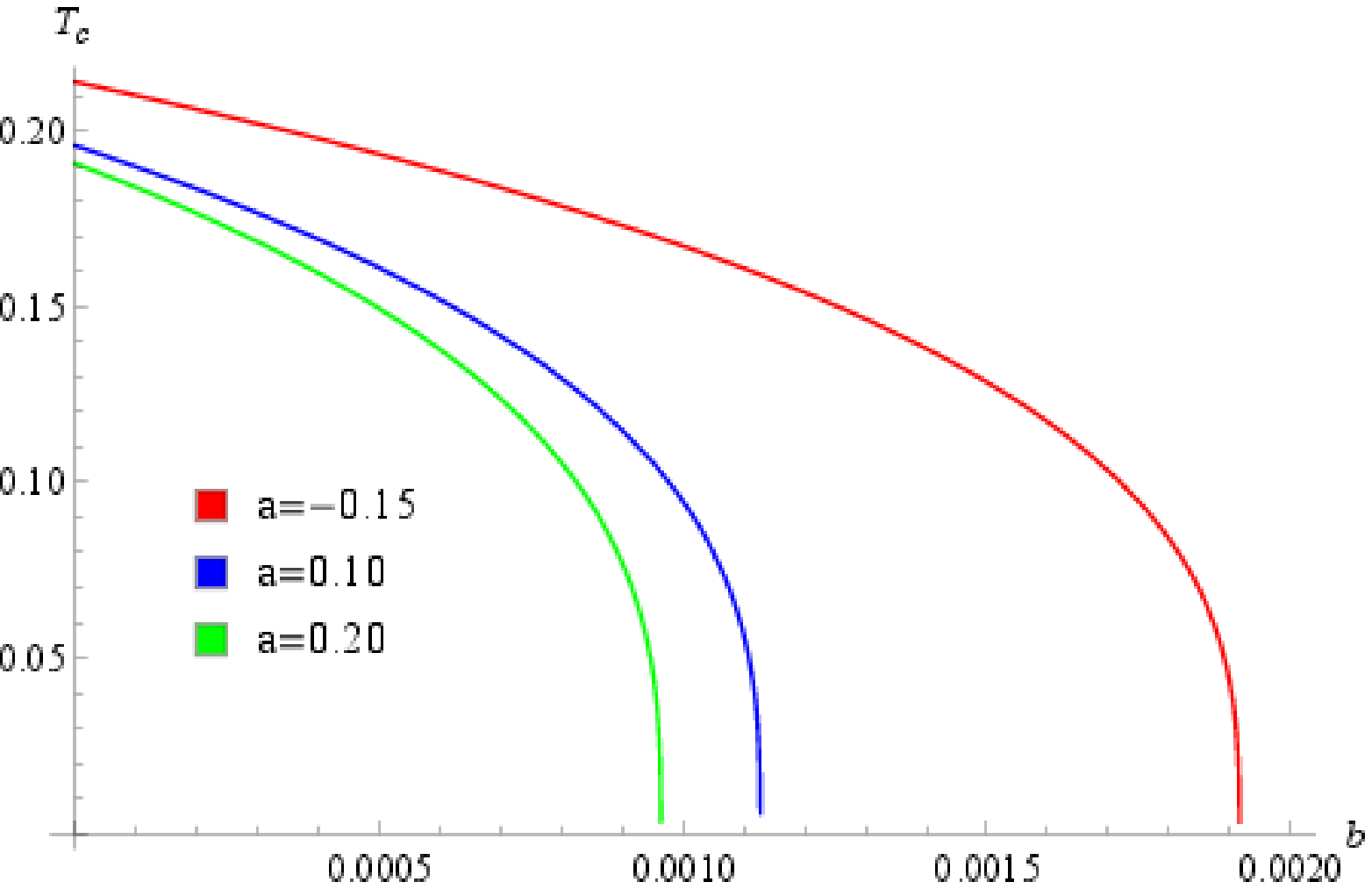}
\vfill
\includegraphics[width=.45\textwidth,trim=0 0 0 0,clip]{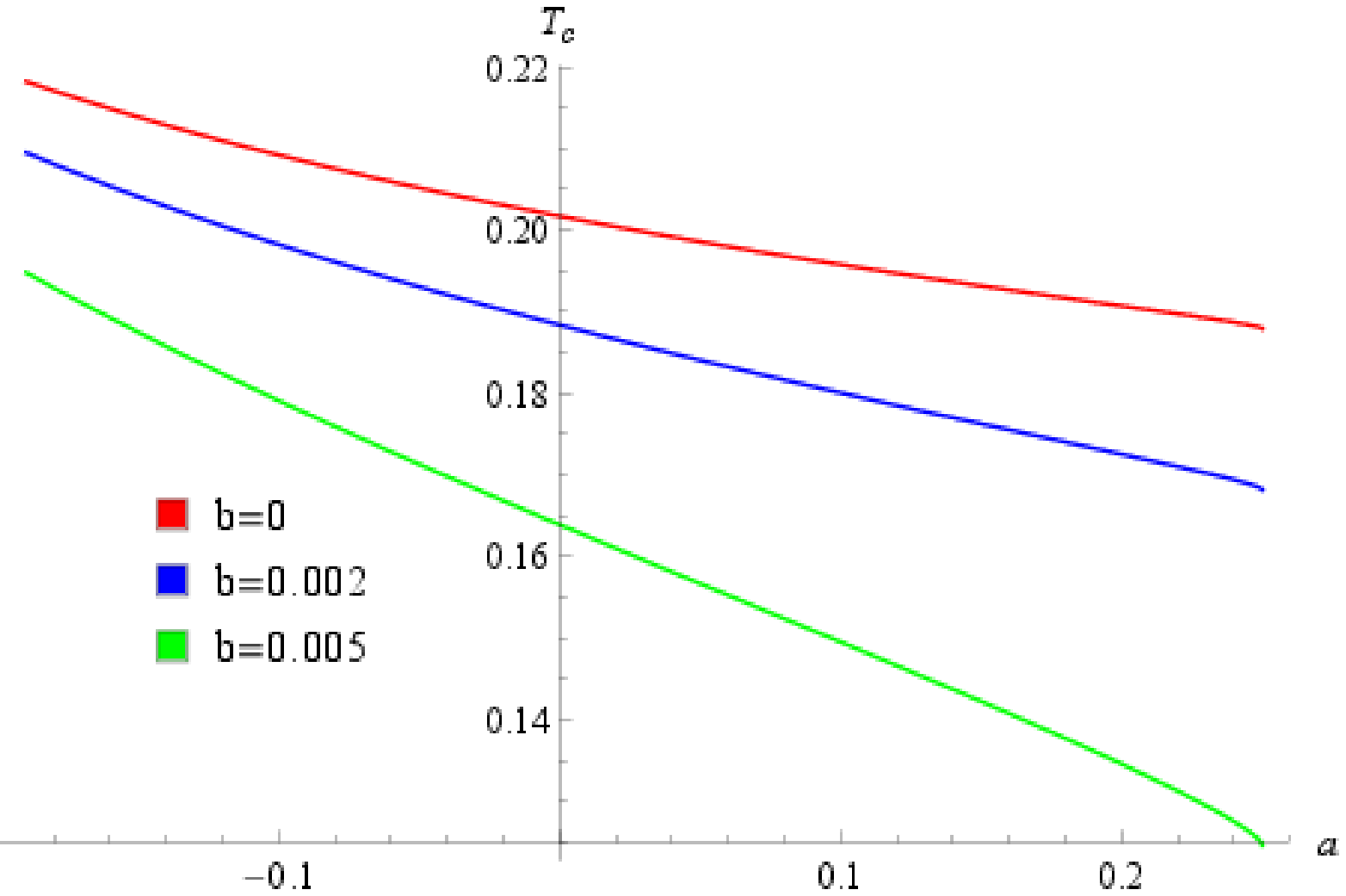}
\hfill
\includegraphics[width=.45\textwidth,origin=c,angle=0,]{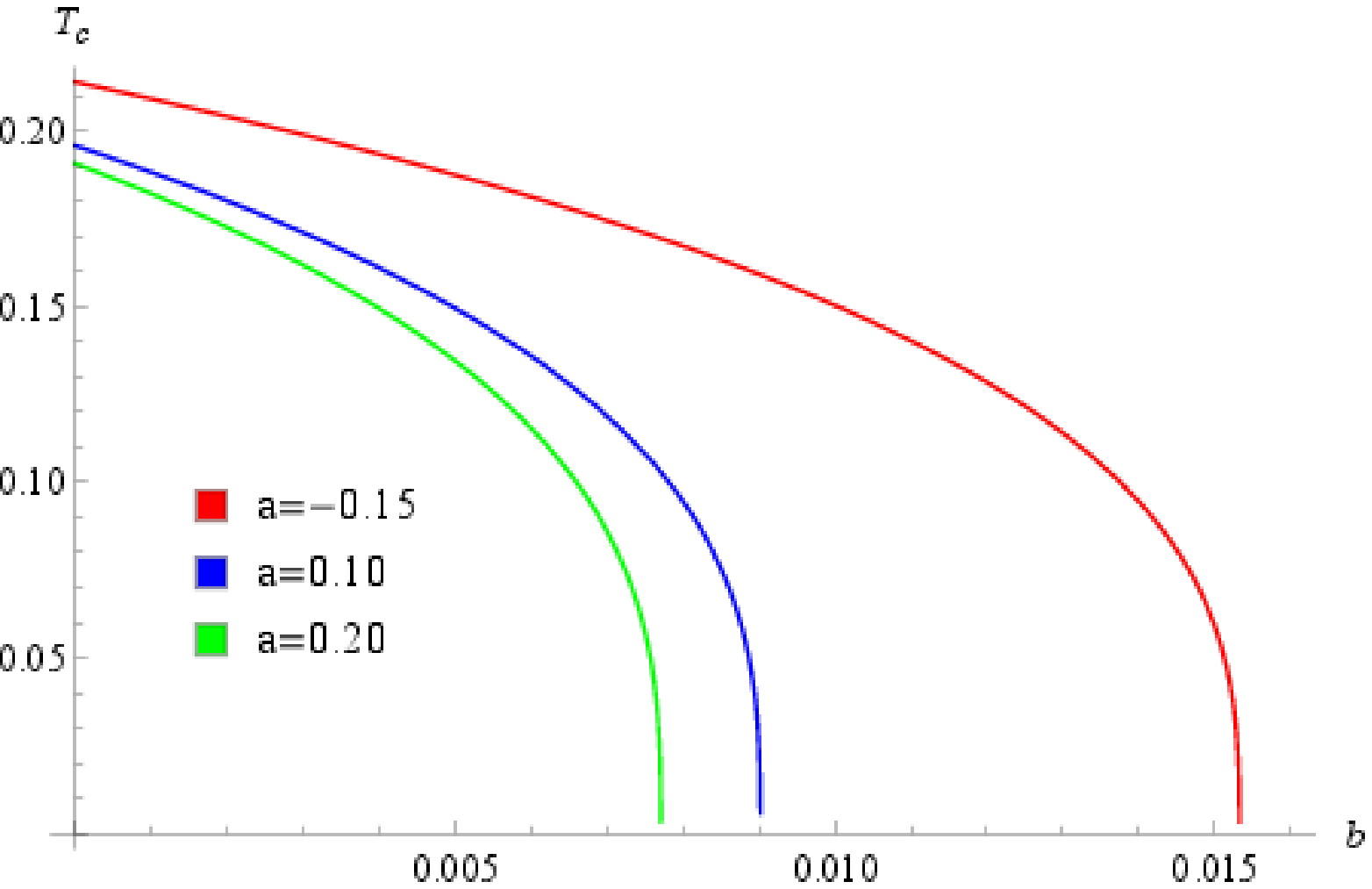}
\caption{\label{fig:2} Left panel: Variation of $T_{c}$ with $a$ for different values of $b$ (top for ENE, bottom for LNE); Right panel: Variation of $T_{c}$ with $b$ for different values of $a$ (top for ENE, bottom for LNE). We have chosen $z_{m}=0.5$ and $\rho=1$. }
\end{figure}

\hspace*{1cm}(iv) The Gauss-Bonnet parameter ($a$) also has important consequences in the formation of the holographic condensate. From Table:\ref{tab:1} it is clear that as we increase the value of $a$ the critical temperature for condensation decreases. This means that the increase of $a$ makes the formation of scalar hair difficult. This indeed shows that both $a$ and $b$ has the same kind of influences on the formation of the hair. However, from Figure:\ref{fig:2} we observe that $T_{c}$ decreases more rapidly with $b$ than with $a$. This clearly suggests that the Born-Infeld parameter ($b$) modifies the critical temperature more significantly than the Gauss-Bonnet parameter ($a$).

\hspace*{1cm}(v) The expectation value of the condensation operator, $\langle\mathcal{O}_{2}\rangle$, vanishes at the critical point $T=T_{c}$ and the condensation occurs below the critical temperature, $T_{c}$ (see the right panel of Figure:\ref{fig:1}). Moreover, form \eqref{eq:3.17} (and \eqref{eq:A2}) we observe that $\langle\mathcal{O}_{2}\rangle\propto(1-T/T_{c})^{1/2}$ which shows the mean field behaviour of the holographic condensates and signifies that there is indeed a second order phase transition (critical exponent $1/2$). This also admires the consistency of our analysis.







 



\begin{table}[tbp]
  \begin{minipage}{0.5\linewidth}
      
\begin{tabular}{@{}|r|rr|rr|@{}}
\toprule
$b$ & \multicolumn{2}{c}{$a = -0.19$} & \multicolumn{2}{c}{$a = -0.10$}  \\
\cmidrule{2-3} \cmidrule{4-5} 
& ENE & LNE & ENE & LNE \\ \midrule

0.0002 & 0.2113 & 0.2174 & 0.2005 & 0.2081\\

0.0004 & 0.2039 & 0.2165 & 0.1910 & 0.2070\\ 

0.0006 & 0.1958 & 0.2157 & 0.1805 & 0.2060 \\ 

0.0008 & 0.1871 & 0.2148 & 0.1686 & 0.2049 \\

0.0010 & 0.1775 & 0.2139 & 0.1547 & 0.2039 \\ 

0.0012 & 0.1667 & 0.2130 & 0.1377 & 0.2028 \\ 
 
0.0014 & 0.1542 & 0.2122 & 0.1149 & 0.2016  \\

0.0016 & 0.1394 & 0.2113 & 0.0753 & 0.2005 \\

0.0018 & 0.1204 & 0.2104 & --- & --- \\

0.0020 & 0.0923 & 0.2094 & --- & --- \\
\bottomrule
        \end{tabular}
    \end{minipage}  
     \begin{minipage}{0.5\linewidth}
       
            \begin{tabular}{@{}|r|rr|rr|@{}}
\toprule
$b$ & \multicolumn{2}{c}{$a = 0.10$} & \multicolumn{2}{c}{$a = 0.25$}  \\
\cmidrule{2-3} \cmidrule{4-5} 
& ENE & LNE & ENE & LNE \\ \midrule

0.0002 & 0.1834 & 0.1943 & 0.1725 & 0.1861\\

0.0003 & 0.1766 & 0.1936 & 0.1636 & 0.1852\\

0.0004 & 0.1692 & 0.1928 & 0.1537 & 0.1843\\ 

0.0005 & 0.1610 & 0.1920 & 0.1421 & 0.1834\\

0.0006 & 0.1520 & 0.1913 & 0.1285 & 0.1824 \\

0.0007 & 0.1417 & 0.1906 & 0.1110 & 0.1815\\ 

0.0008 & 0.1296 & 0.1898 & 0.0852 & 0.1805 \\

0.0009 & 0.1148 & 0.1890 & --- & --- \\

0.0010 & 0.0946 & 0.1882 & --- & --- \\ 

 



\bottomrule
            \end{tabular}
        \end{minipage}    
\caption{{\label{tab:1}}Numerical values of coefficients of $T_{c}$ for different values of the parameters $a$, $b$.}
\end{table}
\label{sec:nomag}

\section{Effects of external magnetic field with non-linear corrections}
In this section we intend to study the effect of an external magnetic field on the holographic superconductors with the non-linear electrodynamics mentioned earlier. But before we present our analysis, we would like to briefly mention the \textit{Meissner-like effect} in the context of holographic superconductors\cite{ref64} which will be central to our discussion. It is observed that when immersed in an external magnetic field, ordinary superconductors expel magnetic field lines thereby exhibiting perfect diamagnetism when the temperature is lowered through $T_{c}$. This is the Meissner effect\cite{ref70}. But for the holographic superconductors in the probe limit we neglect the backreaction of the scalar field on the background geometry. As a result the superconductors are not able to repel the background magnetic field. Instead the scalar condensates adjust themselves such that they only fill a finite strip in the plane which reduces the total magnetic field passing through it. In other words, the effect of the external magnetic field is such that it always tries to reduce the condensate away making the condensation difficult to set in. Considering this apparent similarity with the conventional Meissner effect, this holographic phenomena is referred to as \textit{Meissner-like effect}.

In order to study the effects of magnetic field on the holographic superconductors we add an external static magnetic field in the bulk. According to the gauge/gravity duality, the asymptotic value of the magnetic field in the bulk corresponds to a magnetic field in the boundary field theory, i.e., $B({\bf{x}})= F_{xy}({\bf{x}}, z\rightarrow 0)$\cite{ref64,ref65}. Considering the fact that, near the critical magnetic field, $B_{c}$, the value of the condensate is small, we may consider the scalar field $\psi$ as a perturbation near $B_{c}$. This allows us to adopt the following ansatz for the gauge field and the scalar field\cite{ref54,ref55,ref64,ref65}:
\begin{subequations}\label{eq:4.1}
\begin{align}
\label{eq:4.1:1}
A_{\mu} =&\Big(\phi(z),0,0,Bx,0\Big),
\\
\label{eq:4.1:2}
\psi=&\psi(x,z).
\end{align}
\end{subequations}

With the help of \eqref{eq:2.8}, \eqref{eq:4.1:1} and \eqref{eq:4.1:2} we may write the equation of motion for the scalar field $\psi(x,z)$ as\cite{ref54,ref55,ref63},
\begin{equation}
\label{eq:4.2}
\psi''(x,z)-\frac{\psi'(x,z)}{z} + \frac{f'(z)}{f(z)}\psi'(x,z) + \frac{r_{+}^{2}\phi^{2}(z)\psi(x,z)}{z^{4}f^{2}(z)}+\frac{1}{z^{2} f(z)}\left(\partial_{x}^{2}\psi-B^{2}x^{2}\psi\right)+\frac{3r_{+}^{2}\psi(x,z)}{z^{4}f(z)}=0
\end{equation}

In order to solve \eqref{eq:4.2} we shall use the method of separation of variables\cite{ref64,ref65}. Let us consider the solution of the following form:
\begin{equation}
\label{eq:4.3}
\psi(x,z)=X(x)R(z).
\end{equation}

As a next step, we shall substitute \eqref{eq:4.3} into \eqref{eq:4.2}. This yields the following equation which is separable in the two variables, $x$ and $z$.
\begin{equation}
\label{eq:4.4}
z^{2}f(z)\left[\dfrac{R''(z)}{R(z)}+\dfrac{R'(z)}{R(z)}\left(\dfrac{f'(z)}{f(z)}-\dfrac{1}{z}\right)+\dfrac{r_{+}^{2}\phi^{2}(z)}{z^{4}f^{2}(z)}+\dfrac{3r_{+}^{2}}{z^{4}f(z)}\right]-\left[-\dfrac{X''(x)}{X(x)}+B^{2}x^{2}\right]=0.
\end{equation}

It is interesting to note that, the $x$ dependent part of \eqref{eq:4.4} is localized in one dimension. Moreover, this is exactly solvable since it maps the quantum harmonic oscillator (QHO). This may be identified as the Schr\"{o}dinger equation for the corresponding QHO with a frequency determined by $B$\cite{ref54,ref55,ref64,ref65},
\begin{equation}
\label{eq:4.5}
-X''(x)+B^{2}x^{2}X(x)=C_{n}BX(x)
\end{equation}
where $C_{n}=2n+1$ ($n=integer$). Since the most stable solution corresponds to $n=0$\cite{ref54,ref55,ref64}, the $z$ dependent part of \eqref{eq:4.4} may be expressed as
\begin{equation}
\label{eq:4.6}
R''(z)+ \left(\frac{f'(z)}{f(z)}-\frac{1}{z}\right)R'(z)+\frac{r_{+}^{2}\phi^{2}(z)R(z)}{z^{4}f^{2}(z)}+\frac{3r_{+}^{2} R(z)}{z^{4}f(z)}=\frac{BR(z)}{z^{2}f(z)}.
\end{equation}

Now at the horizon, $z=1$, using \eqref{eq:2.15} and \eqref{eq:4.6}, we may write the following equation:
\begin{equation}
\label{eq:4.7}
R'(1)=\left(\frac{3}{4}-\frac{B}{4r_{+}^{2}}\right)R(1).
\end{equation}

On the other hand, at the asymptotic infinity, $z\rightarrow 0$, the solution of \eqref{eq:4.6} can be written as
\begin{equation}
\label{eq:4.8}
R(z)=D_{-}z^{\lambda_{-}}+D_{+}z^{\lambda_{+}}.
\end{equation}

It is to be noted that in our analysis we shall choose $D_{-}=0$ as was done in section~\ref{sec:nomag}.

Near the horizon, $z=1$, Taylor expansion of $R(z)$ gives
\begin{equation}
\label{eq:4.9}
R(z)= R(1)-R'(1)(1-z)+\frac{1}{2}R''(1)(1-z)^{2} +\cdots
\end{equation}
where we have considered $R'(1)<0$ without loss of generality.

Now calculating $R''(1)$ from \eqref{eq:4.6} and using \eqref{eq:4.7} we may write from \eqref{eq:4.9}
\begin{equation}
\label{eq:4.10}
R(z)=\frac{1}{4}R(1)+\frac{3z}{4}R(1)+(1-z)\frac{B}{4r_{+}^{2}}R(1) 
+\frac{1}{2}(1-z)^{2}\left[3a-\frac{15}{32}+(1-16a)\frac{B}{16r_{+}^{2}} +\frac{B^{2}}{32r_{+}^{4}} -\frac{\phi'^{2}(1)}{32r_{+}^{2}}\right]R(1)
\end{equation}
where in the intermediate step we have used the Leibniz rule [cf.~\eqref{eq:3.3}].

Finally, matching the solutions \eqref{eq:4.8} and \eqref{eq:4.10} at the intermediate point $z=z_{m}$ and performing some simple algebraic steps as in section~\ref{sec:nomag} we arrive at the following equation in $B$:
\begin{eqnarray}
\label{eq:4.11}
B^{2} + 2Br_{+}^{2}\Bigg[\frac{8\left(\lambda_{+}-(\lambda_{+}-1)z_{m}\right)}{(1-z_{m})(\lambda_{+}-\lambda_{+}z_{m}+2z_{m})}+(1-16a)\Bigg]
+\Bigg[\frac{(1+3z_{m})
\lambda_{+}-3z_{m}}{2(1-z_{m})(\lambda_{+}-\lambda_{+}z_{m}+2z_{m})} \\ \nonumber
+\left(3a-\frac{15}{32}-\frac{\phi'^{2}(1)}{32r_{+}^{2}}\right)\Bigg]32r_{+}^{4}=0.
\end{eqnarray}

Equation~\eqref{eq:4.11} is quadratic in $B$ and its solution is found to be of the following form\cite{ref63}:
\begin{eqnarray}
\label{eq:4.12}
B&=&r_{+}^{2}\Bigg[\Bigg(\frac{8\left(\lambda_{+}-(\lambda_{+}-1)z_{m}\right)}{(1-z_{m})(\lambda_{+}-\lambda_{+}z_{m}+2z_{m})}+(1-16a)\Bigg)^{2}-\Bigg(\frac{16\left[
(1+3z_{m})
\lambda_{+}-3z_{m}\right]}{(1-z_{m})(\lambda_{+}-\lambda_{+}z_{m}+2z_{m})} \\ \nonumber
&&+\left(96a-15-\frac{\phi'^{2}(1)}{r_{+}^{2}}\right)\Bigg)\Bigg]^{\dfrac{1}{2}}
-r_{+}^{2}\Bigg(\frac{8\left(\lambda_{+}-(\lambda_{+}-1)z_{m}\right)}{(1-z_{m})(\lambda_{+}-\lambda_{+}z_{m}+2z_{m})}+(1-16a)\Bigg).
\end{eqnarray}

We are interested in determining the critical value of the magnetic field strength, $B_{c}$, above which the superconducting phase disappears. In this regard, we would like to consider the case for which $B\sim B_{c}$. Interestingly, in this case the condensation becomes vanishingly small and we can neglect terms that are quadratic in $\psi$. Thus, the equation of motion corresponding to the gauge field (equation~\eqref{eq:2.13}), $\phi$, may be written as
\begin{equation}
\label{eq:4.13}
\Bigg(1+\dfrac{4bz^{4}\phi'^{2}(z)}{r_{+}^{2}}\Bigg)\phi''(z) -\frac{1}{z}\phi'(z)+\dfrac{8bz^3}{r_+^2}\phi'^{3}(z)=0.
\end{equation}

In order to solve the above equation we shall consider a perturbative solution of the following form:
\begin{equation}
\label{eq:4.14}
\phi(z)=\phi_{0}(z)+\frac{b}{r_{+}^{2}}\phi_{1}(z)+\cdots
\end{equation}
where $\phi_{0}(z)$ is the solution of $\phi(z)$ for $b=0$ and so on. After some algebraic calculations we may write the solution of \eqref{eq:4.14} as\footnote{See Appendix A for the derivation.}
\begin{equation}
\label{eq:4.15}
\phi(z)=\dfrac{\rho}{r_{+}^{2}}(1-z^{2})\left[1+\frac{b}{r_{+}^{2}}-\frac{2b\rho^{2}}{r_{+}^{6}}(1+z^{4})(1+z^{2})\right].
\end{equation}

At this point of discussion, it must be stressed that we have considered terms which are linear in the non-linear parameter $b$.

At the asymptotic boundary of the AdS space, $z=0$, the solution \eqref{eq:4.15} can be approximated as
\begin{equation}
\label{eq:4.16}
\phi(z)\approx\dfrac{\rho}{r_{+}^{2}}\left[1+\frac{b}{r_{+}^{2}}-\frac{2b\rho^{2}}{r_{+}^{6}}\right]-\dfrac{\rho}{r_{+}^{2}}\left(1 + \dfrac{b}{r_{+}^{2}}\right)z^{2}.
\end{equation}

Now, comparing \eqref{eq:4.16} with \eqref{eq:2.16:1} we may identify the chemical potential, $\mu$, as
\begin{equation}
\label{eq:4.17}
\mu = \dfrac{\rho}{r_{+}^{2}}\left[1+\frac{b}{r_{+}^{2}}-\frac{2b\rho^{2}}{r_{+}^{6}}\right]
\end{equation}

Near the horizon, $z=1$, we may write from \eqref{eq:4.13}
\begin{equation}
\label{eq:4.18}
\phi''(1)=\phi'(1)-\dfrac{12b}{r_{+}^{2}}\phi'^{3}(1)+ \mathcal{O}(b^{2})
\end{equation}

Substituting \eqref{eq:4.18} into \eqref{eq:3.1} and using the boundary condition \eqref{eq:2.15} we may write
\begin{equation}
\label{eq:4.19}
\phi(z)=-\phi'(1)(1-z)+\frac{1}{2} (1-z)^{2}\left(\phi'(1)-\dfrac{12b}{r_{+}^{2}}\phi'^{3}(1)\right)
\end{equation}

Matching the solutions \eqref{eq:4.19} and \eqref{eq:2.16:1} at the intermediate point $z_{m}$ and using \eqref{eq:4.17} we can find the following relation:
\begin{equation}
\label{eq:4.20}
(\beta-2\eta)(2\eta^{3}-6\beta^{3}-\eta)=0
\end{equation} 
where we have set $-\phi'(1)=\beta$ and $\frac{\rho}{r_{+}^{2}}=\eta$. One of the solutions of this quartic equation can be written as
\begin{equation*}
\label{eq:4.21}
\beta=2\eta
\end{equation*}
which implies
\begin{equation}
\label{eq:4.22}
\phi'(1)=-\dfrac{2\rho}{r_{+}^{2}}.
\end{equation}

As a final step, substituting \eqref{eq:4.22} into \eqref{eq:4.12} and using \eqref{eq:2.5} and \eqref{eq:3.14} we obtain the critical value of the magnetic field strength as
\begin{equation}
\label{eq:4.23}
\dfrac{B_{c}}{T_{c}^{2}}=\pi^{2}\left(1 + \frac{12b\tilde{\beta}^{2}(1-z_{m})}{\mathcal{C}^{2}z_{m}}\right)\left[\tilde{\beta}\mathcal{C}-\mathcal{M}\left(\dfrac{T}{T_{c}}\right)^{3}\right]
\end{equation} 

In a similar manner we can determine the critical value of magnetic strength, $B_{c}$, for the holographic superconductor with logarithmic electrodynamics. The expression for $B_{c}$ in this model is given in Appendix~\ref{LNE}.

In \eqref{eq:4.23} the terms $\mathcal{C}$ and $\mathcal{M}$ can be identified as
\begin{equation*}
\mathcal{C}=\left(1-\frac{\mathcal{A}-\mathcal{M}^{2}}{\tilde{\beta}^{2}}x^{6}\right)^{\frac{1}{2}}
\end{equation*}
\begin{equation*}
\mathcal{M}=1-16a+\frac{8(\lambda_{+}-(\lambda_{+}-1)z_{m})}{(1-z_{m})(\lambda_{+}-\lambda_{+}z_{m}+2z_{m})}
\end{equation*}
where
\begin{equation*}
\mathcal{A}=\left[96a-15+16\left(\frac{(1+3z_{m})\lambda_{+}-3z_{m}}{(1-z_{m})(\lambda_{+}-\lambda_{+}z_{m}+2z_{m})}\right)\right].
\end{equation*}

Here, we must mention that we have normalized $B_{c}$ by the square of the critical temperature, $T_{c}$, such that the critical magnetic field strength, $B_{c}$, becomes dimensionless.

\begin{figure}[tbp]
\centering 
\includegraphics[width=.45\textwidth,trim=0 0 0 0,clip]{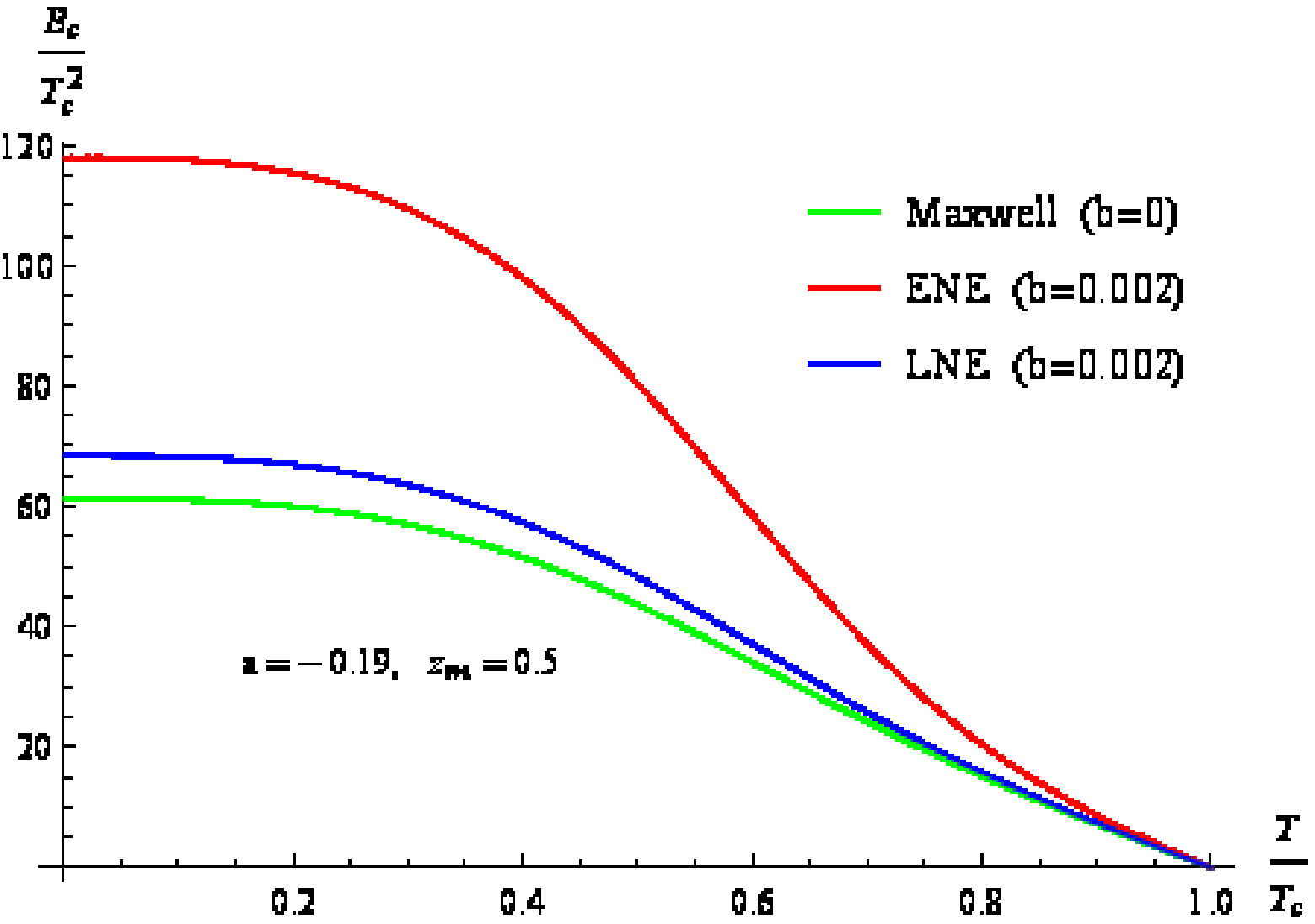}
\hfill
\includegraphics[width=.45\textwidth,origin=c,angle=0,]{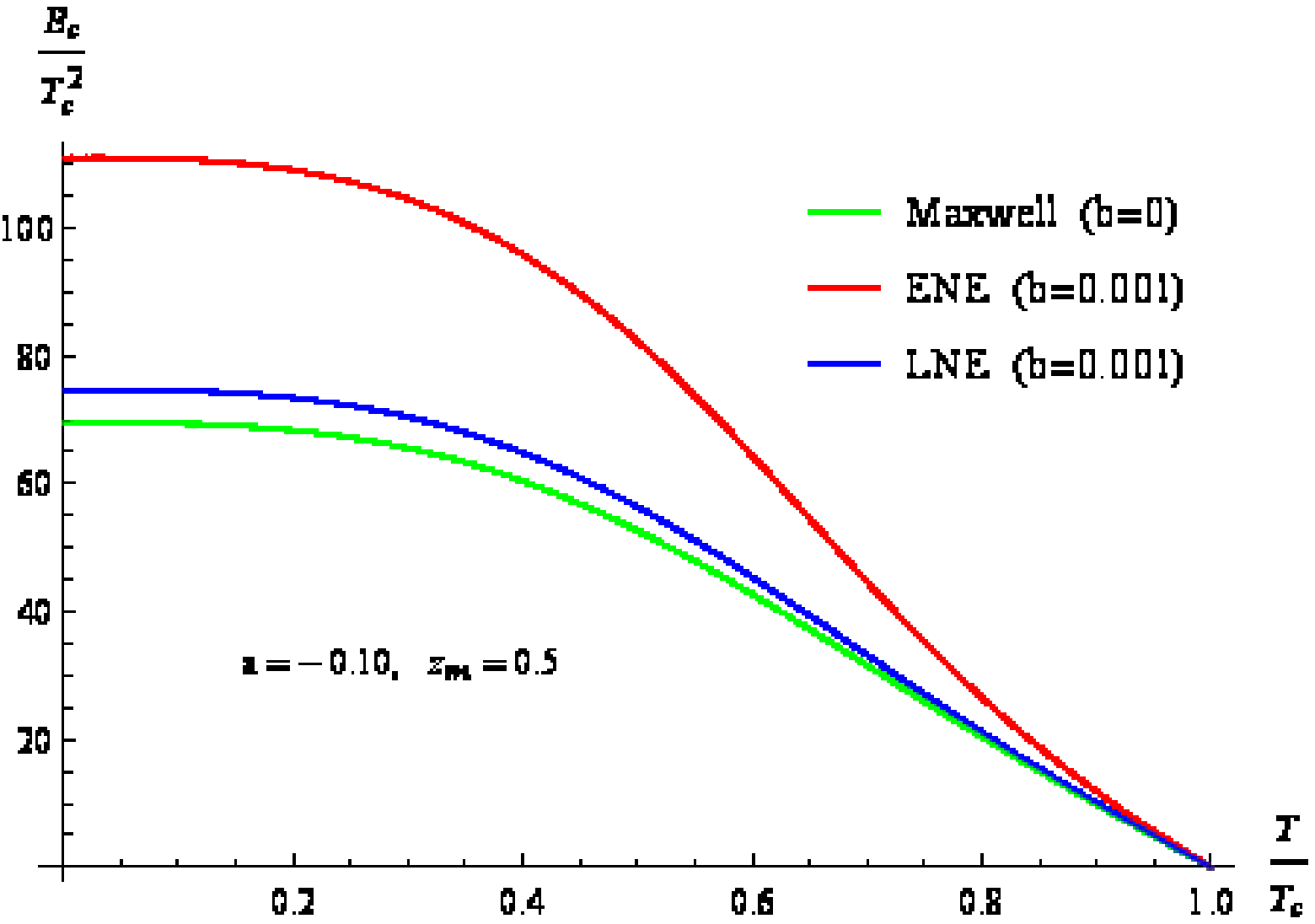}
\vfill
\includegraphics[width=.45\textwidth,trim=0 0 0 0,clip]{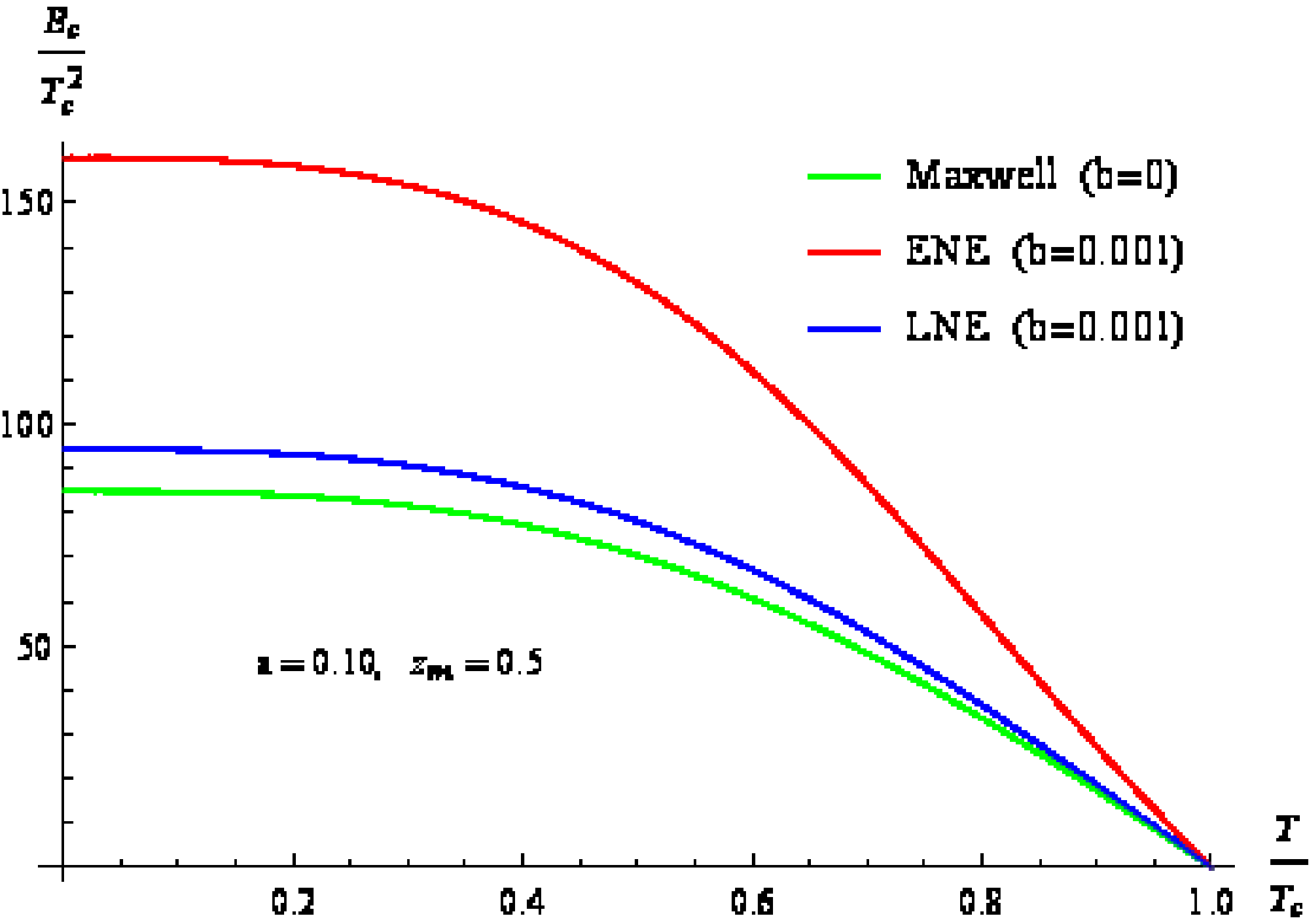}
\hfill
\includegraphics[width=.45\textwidth,trim=0 0 0 0,clip]{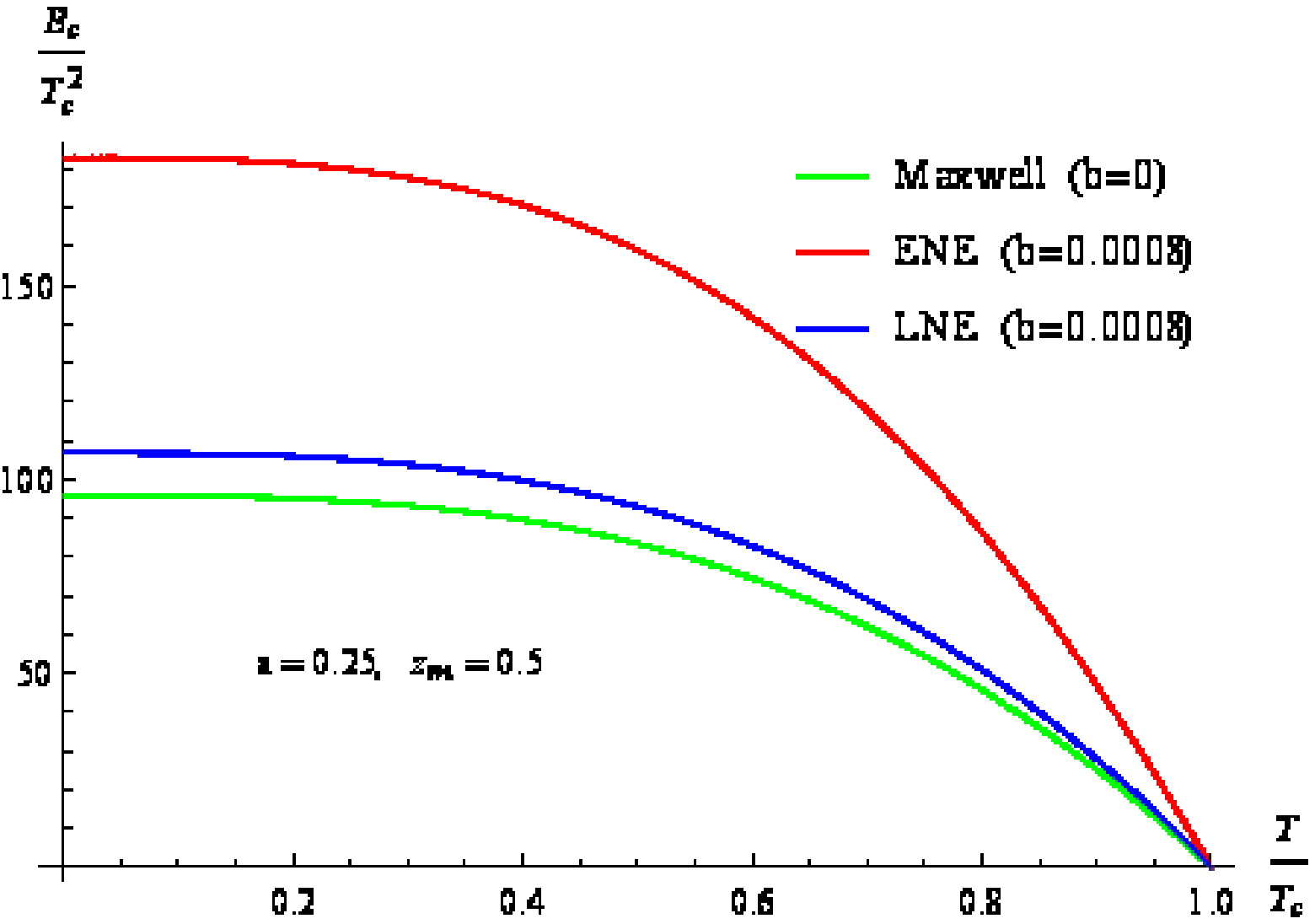}
\caption{\label{fig:3} Plots of $B_{c}/T_{c}^{2}-T/T_{c}$ for different values of $a$ and $b$.}
\end{figure}

In figure~\ref{fig:3} we have plotted \eqref{eq:4.23} (and \eqref{eq:A3}) as a function of $T/T_{c}$. From these plots it is evident that above the critical magnetic field ($B_{c}$) the superconductivity is completely destroyed which is also the case for ordinary type II superconductors\cite{ref70}.  From the above analysis we can explain the effects of the Gauss-Bonnet coupling parameter ($a$) and the non-linear parameter ($b$) on the holographic condensates. First of all we note that, for fixed values of $a$ the critical magnetic field ($B_{c}$) increases with $b$. Secondly, the critical magnetic field corresponding to the Maxwell case ($b=0$) is lower than those for ENE and LNE. This indicates that the critical magnetic field strength is higher in presence of the non-linear corrections than the usual Maxwell case. Moreover, this increment is larger for the holographic condensate with ENE than that with LNE. Finally, if we vary $a$ while keeping $b$ constant similar effects are observed, i.e. the critical field strength increases with the Gauss-Bonnet parameter ($a$). From the preceding discussion we may infer that both the higher order corrections indeed make the condensation harder to form. Moreover, between the two non-linear electrodynamics, the exponential electrodynamics has stronger effects on the formation of the holographic $s$-wave condensate namely, the formation of the scalar hair is more difficult for holographic superconductor with ENE. This can be explained by noting that the increase in the critical field strength ($B_{c}$) tries to reduce the condensate away making the condensation difficult to set in\cite{ref54,ref55,ref64}. 
\label{sec:mag}
\section{Conclusions}
In this paper, considering the probe limit, we have studied a holographic model of superconductor in the higher curvature planar Gauss-Bonnet-AdS black hole background. We have also taken into account two different types of non-linear electrodynamics (exponential and logarithmic non-linear electrodynamics) in the matter Lagrangian which may be considered as higher derivative corrections to the gauge fields in the usual Abelian gauge theory. In addition to that, we have made an analytic investigation on the effects of an external magnetic field on these superconductors.
The primary motivations of the present paper is to study the effects of several non-linear corrections to the gravity and matter sectors of the action (that describes the holographic superconductivity) on the holographic $s$-wave condensates both in presence as well as absence of an external magnetic field. Along with this we aim to make a comparative study among the usual Maxwell electrodynamics and the two NEDs considered in the paper (ENE, LNE) regarding their effects on the formation of holographic condensates. Based on purely analytic methods we have successfully addressed these issues. The main results of our analysis can be put as follows:
\begin{itemize}

\item Non-linear electrodynamics has stronger effects on the condensates than the usual Maxwell case. The critical temperature for condensation ($T_{c}$) decreases as we increase the values of the non-linear parameter ($b$) as well as the Gauss-Bonnet coupling parameter ($a$) (Figure:\ref{fig:1}, Table:\ref{tab:1}). Moreover, $b$ modifies the critical temperature more significantly than $a$ (Figure:\ref{fig:2}). On the other hand, the normalized order parameter ($\langle\mathcal{O}_{2}\rangle^{1/\lambda_{+}}/T_{c}$) increases with the increase of $b$ and $a$ (Figure:\ref{fig:1}). This implies that, in the presence of the higher order corrections the formation of the scalar hairs become difficult. 

\item The variation of the order parameter with temperature, $\langle\mathcal{O}_{2}\rangle\propto(1-T/T_{c})^{1/2}$, exhibits a mean-field behavior. Also, the value of the associated critical exponent is $1/2$, which further ensures that the holographic condensates indeed undergo a second order phase transition in going from normal to superconducting phase. 

\item There exists a critical magnetic field $B_{c}$ above which the superconductivity ceases to exist (Figure:\ref{fig:3}). This property is similar to that of ordinary type II superconductors\cite{ref70}. Also, the critical magnetic field strength ($B_{c}$) increases as we increase both $b$ and $a$. The increasing magnetic field strength tries to reduce the condensate away completely making the condensation difficult to form.

\item From Figures:\ref{fig:1},\ref{fig:3} and Table:\ref{tab:1} we further observe that for particular parameter values $T_{c}$ is less in holographic superconductor with ENE than that with LNE whereas, $\langle\mathcal{O}_{2}\rangle^{1/\lambda_{+}}/T_{c}$ and $B_{c}$ is more in the previous one. These results suggest that the exponential electrodynamics exhibit stronger effects than the logarithmic electrodynamics.

\end{itemize}


It is interesting to note that similar conclusions were drawn in Ref.\cite{ref59} where numerical computations were performed in this direction. Our analytic calculations provide further confirmations regarding this issue. However, the novel feature of our present analysis is that we have been able to study the effect of the higher curvature corrections which was not performed explicitly in Ref.\cite{ref59}.

Although, we have been able to explore several issues regarding $s$-wave holographic superconductors with different non-linear corrections, there are several other nontrivial and important aspects which can be explored in the future. These can be written in the following order:

\hspace*{1cm}(i) We have performed our entire analysis in the probe limit. It will be interesting to carry out the analysis by considering the back-reaction on the metric.

\hspace*{1cm}(ii) We can further extend our analysis considering higher curvature gravity theories beyond Gauss-Bonnet gravity. Also, the study of holographic $p$-wave and $d$-wave superconductors in presence of these various higher order corrections may be taken into account.
\section{Acknowledgements}
The authors would like to thank Rabin Banerjee and Dibakar Roychowdhury for useful discussions. A. L would like to thank Council of Scientific and Industrial Research (C. S. I. R), Government of India, for financial support.
\appendix
\numberwithin{equation}{section}
\section{Appendix}
\subsection{Holographic condensate with LNE:}\label{LNE}

\hspace{0.6cm}The critical temperature for condensation:
\begin{equation}
\label{eq:A1}
T_{c}=\left[\dfrac{2\rho}{\tilde{\beta}\pi^{3}}\left(1-\frac{3b\tilde{\beta}^{2}(1-z_{m})}{2z_{m}}\right)\right]^{\frac{1}{3}}.
\end{equation}

Normalized order parameter:
\begin{equation}
\label{eq:A2}
\dfrac{\langle
\mathcal{O}_{2}\rangle^{\frac{1}{\lambda_{+}}}}{T_{c}} =\left(\frac{\pi}{z_{m}}\right)\left[\dfrac{(3z_{m}^{2}+5z_{m})}{4\left(\lambda_{+}+(2-\lambda_{+})z_{m}\right)}\right]^{\frac{1}{\lambda_{+}}}
\left[\dfrac{6z_{m}}{(1-z_{m})}\left(1+\frac{3b\tilde{\beta}^{2}(2-z_{m})}{4z_{m}}\right)\left(1-\dfrac{T}{T_{c}}\right)\right]^{\frac{1}{2\lambda_{+}}}.
\end{equation}

Critical magnetic field strength:
\begin{equation}
\label{eq:A3}
\dfrac{B_{c}}{T_{c}^{2}}=\pi^{2}\left(1 + \frac{3b\tilde{\beta}^{2}(1-z_{m})}{2\mathcal{C}^{2}z_{m}}\right)\left[\tilde{\beta}\mathcal{C}-\mathcal{M}\left(\dfrac{T}{T_{c}}\right)^{3}\right].
\end{equation}

The quantities $\mathcal{C},~\mathcal{M}$ and $\mathcal{A}$ are identified in section~\ref{sec:mag}.

\subsection{\textbf{Solution of gauge field ($\phi(z)$) near $B_{c}$ :}}\label{PHI}
The equation for the gauge field near $B_{c}$ for the holographic superconductor with ENE is given by
\begin{equation}
\label{eq:A.1}
\Bigg(1+\dfrac{4bz^{4}\phi'^{2}(z)}{r_{+}^{2}}\Bigg)\phi''(z) -\frac{1}{z}\phi'(z)+\dfrac{8bz^3}{r_+^2}\phi'^{3}(z)=0.
\end{equation}

Let us consider the following perturbative solution of \eqref{eq:A.1}:
\begin{equation}
\label{eq:A.2}
\phi(z)=\phi_{0}(z)+\frac{b}{r_{+}^{2}}\phi_{1}(z)+\cdots
\end{equation}
where $\phi_{0}(z)$, $\phi_{1}(z)$, $\cdots$ are independent solutions and the numbers in the suffices of $\phi(z)$ indicate the corresponding order of the non-linear parameter ($b$).

Substituting \eqref{eq:A.2} in \eqref{eq:A.1} we obtain
\begin{equation}
\label{eq:A.3}
\left[\phi_{0}''(z)-\dfrac{\phi_{0}'(z)}{z}\right]+\dfrac{b}{r_{+}^{2}}\left[\phi_{1}''(z)-\dfrac{\phi_{1}'(z)}{z}+4z^{4}\phi_{0}''(z)\phi_{0}'^{2}(z)+\dfrac{8z^{3}}{r_{+}^{2}}\phi_{0}'^{3}(z)\right]+\mathcal{O}(b^{2})=0.
\end{equation}

Equating the coefficients of $b^{0}$ and $b^{1}$ from the l.h.s of \eqref{eq:A.3} to zero we may write
\begin{subequations}\label{eq:A.4}
\begin{align}
\label{eq:A.4:1}
b^{0}:&\qquad\phi_{0}''(z)-\dfrac{\phi_{0}'(z)}{z}=0
\\
\label{eq:A.4:2}
b^{1}:&\qquad\phi_{1}''(z)-\dfrac{\phi_{1}'(z)}{z}+4z^{4}\phi_{0}''(z)\phi_{0}'^{2}(z)+\dfrac{8z^{3}}{r_{+}^{2}}\phi_{0}'^{3}(z)=0
\end{align}
\end{subequations}

Now, using the boundary condition \eqref{eq:2.16:1} we may write the solution of \eqref{eq:A.4:1} as
\begin{equation}
\label{eq:A.5}
\phi_{0}(z)=\dfrac{\rho}{r_{+}^{2}}(1-z^{2}).
\end{equation}

Using \eqref{eq:A.5} we may simplify \eqref{eq:A.4:2} as
\begin{equation}
\label{eq:A.6}
\phi_{1}''(z)-\dfrac{\phi_{1}'(z)}{z}-96z^{6}\left(\dfrac{\rho}{r_{+}^{2}}\right)^{3}=0.
\end{equation}

As a next step, using the asymptotic boundary condition \eqref{eq:2.16:1}, from \eqref{eq:A.6} we obtain the solution of $\phi_{1}(z)$ as
\begin{equation}
\label{eq:A.7}
\phi_{1}(z)=\dfrac{2\rho^{3}}{r_{+}^{6}}(z^{8}-1)-\dfrac{\rho}{r_{+}^{2}}(z^{2}-1).
\end{equation}

Substituting \eqref{eq:A.5} and \eqref{eq:A.7} in \eqref{eq:A.2} we finally obtain the solution of the gauge field as
\begin{equation}
\label{eq:A.8}
\phi(z)=\dfrac{\rho}{r_{+}^{2}}(1-z^{2})\left[1+\frac{b}{r_{+}^{2}}-\frac{2b\rho^{2}}{r_{+}^{6}}(1+z^{4})(1+z^{2})\right].
\end{equation}

The solution of the gauge field for the holographic superconductor with LNE may be obtained by similar procedure and is given below:
\begin{equation}
\label{eq:A.9}
\phi(z)=\dfrac{\rho}{r_{+}^{2}}(1-z^{2})\left[1+\frac{b}{r_{+}^{2}}-\frac{b\rho^{2}}{4r_{+}^{6}}(1+z^{4})(1+z^{2})\right].
\end{equation}

\end{document}